\documentstyle[aps,preprint,tighten,psfig]{revtex}

\begin{document}
%\psdraft  % only draws the frames for PS figures. for testing.
\draft
%\preprint{\vbox{ \it Submitted to Phys. Rev. C  \hfill\rm JLAB-TH-98-xx}}

\title{Quasifree Kaon Photoproduction on Nuclei}
\author{F. X. Lee~$^{1,2}$, T. Mart~$^{1,3}$, C. Bennhold~$^1$, 
        H. Haberzettl~$^1$, L. E. Wright~$^4$}
\address{$^1$~Center for Nuclear Studies, Department of Physics,
     The George Washington University,  Washington, DC 20052, USA \\
$^2$~Jefferson Lab, 12000 Jefferson Avenue, Newport News, VA 23606, USA\\
$^3$~Jurusan Fisika, FMIPA, Universitas Indonesia, Depok 16424, Indonesia \\
$^4$~Institute of Nuclear and Particle Physics,
     Department of Physics and Astronomy,  \\Ohio University,  Athens, 
     OH 45701, USA}
\maketitle

\begin{abstract}
Investigations of the quasifree reaction A$(\gamma, K Y)$B are 
presented in the distorted wave impulse approximation (DWIA).
For this purpose, we present
a revised tree-level model of elementary kaon photoproduction that 
incorporates hadronic form factors consistent with gauge invariance,
uses SU(3) values for the Born couplings and
uses resonances consistent with multi-channel analyses. 
The potential of exclusive quasifree kaon 
photoproduction on nuclei to reveal 
details of the hyperon-nucleus interaction is examined.
Detailed predictions for the coincidence cross section, the photon
asymmetry, and the hyperon polarization and their sensitivities to the 
ingredients of the model are obtained for all six 
production channels.  Under selected kinematics
these observables are found to be sensitive to 
the hyperon-nucleus final state interaction.
Some polarization observables are found to be insensitive
to distortion effects, making them ideal tools 
to search for possible medium modifications of the elementary amplitude.
\end{abstract}
\vspace{1cm}
 % Polarization in interactions and scattering

\section{Introduction}
\label{intro}

With the start of experimental activities at Jefferson Lab
 and other continuous beam electron accelerators 
with sufficient energy and intensity, explorations in hypernuclear
physics through electromagnetic probes are becoming a reality.
The use of kaon photoproduction to excite discrete 
hypernuclear states through the reaction A$(\gamma,K)_Y$B has been 
investigated extensively~\cite{Bennhold89,Cotanch86,Rosenthal88,Cohen89}.
This reaction involves high momentum transfers to the residual nucleus,
resulting in a cross section that is suppressed by nuclear form factors,
and sensitive to the details of hypernuclear transition densities.
The probability of forming such bound states is in fact rather 
small. It was estimated~\cite{Cotanch86} that this formation probability
is around 5-10\% of the total $(\gamma, K^+)$ strength
on nuclear targets. Thus, most of the kaon production events
will come from quasifree production.  

In this work, we present theoretical predictions for 
exclusive quasifree kaon photoproduction, 
A$(\gamma, K Y)$B, in a Distorted Wave Impulse Approximation (DWIA) framework.
This reaction allows for the study of the production
process in the nuclear medium as well as final state interaction (FSI)
effects without being obscured by the details of the nuclear transition.
This is due mainly to the quasifree nature
of the reaction which permits the kinematic flexibility to have small
momentum transfers. Conceptually, the initial nucleus is a target holder 
which presents a bound nucleon to the incoming photon beam.  
The basic reaction $N(\gamma, K Y)$ takes place in the nuclear medium 
producing a continuum kaon and hyperon which interact with the residual 
nucleus as they exit the target. 

The purpose of the present study is two fold:
First, we want to examine the sensitivity of various
observables to the hyperon-nucleus final-state interaction.
The study of the Y-nucleus potential permits access to the
$YN$ interaction, which is much less well-known than the $\pi N$ and 
$NN$ interactions.  This is mainly due to lack of hyperon beams in 
accelerator experiments.  Recently, effective field theories (EFTs) have been
successfully applied to the strong two-nucleon sector~\cite{kaplan98}.  
While the use of EFTs in the SU(2) regime is now well-established, their
range of applicability to SU(3) is much less certain due to the much
larger degree of SU(3) symmetry breaking.  In order to assess the validity of
EFTs in SU(3) a good phenomenological understanding of the $YN$ force is required.
At present, much of our knowledge on the $YN$ interaction is based on 
studies of hypernuclei formed in hadronic reactions such as 
$(K^-,\pi^-)$ and $(\pi^+, K^+)$~\cite{Bertini81,Chrien88}.

The second goal of this study is to establish the kinematic range
within which polarization observables are insensitive
to distortion effects.  This would allow a clearer signal
for possible medium modifications of the elementary operator
to emerge, as suggested in Ref.~\cite{piekarewicz99}.  
One aspect of this investigation is the puzzle of the
"damped resonances" in the second and third resonance regions
as seen in inclusive photoabsorption cross section data on various
nuclei~\cite{Bianchi93}.  The data show an unexpected damping behavior
of the higher resonances when compared with the same process on the 
proton and the deuteron.  In order to isolate the mechanism for 
this mysterious phenomenon the individual exclusive channels need to be investigated.
In Ref.~\cite{piekarewicz99}, the authors - using PWIA - demonstrate
the sensitivity of polarization observables to the elementary amplitude,
while on the other hand they find these observables to be insensitive to
relativistic effects or the specific nuclear target. In this study,
we compare DWIA with PWIA calculations over a wide kinematic range and thus establish the
range of validity for the conclusions drawn in Ref.~\cite{piekarewicz99}.
Experimentally, there is already an approved Jefferson Lab proposal for
Hall B~\cite{Hyde91} to measure this reaction.

The key ingredients in quasifree kaon photoproduction on nuclei are: 
a) the single-particle wave function of the initial nucleon and spectroscopic 
factor,  usually taken from electron scattering,
b) the elementary kaon photoproduction amplitude, obtained from
models of the free processes, 
c) the distorted kaon wave function, and finally, 
d) the hyperon-nucleus final state interaction.  
This framework has been applied in our previous works for pion photo- 
and electro-production~\cite{Lee93} and eta 
photoproduction~\cite{Lee96} from nuclei, and was found to give 
a good description of the experimental data.
This previous agreement with experiment partially justifies the impulse 
approximation implicit in the model outlined above that we will use in our 
analysis.  Preliminary results in this work have been presented 
in a conference talk~\cite{Bennhold97}. 
 
This paper is organized as follows. In Sec.~\ref{sec:elementary}
we discuss the revised elementary operator and compare it to the
currently available experimental data of 
kaon photoproduction on the nucleon. 
Section~\ref{dwia} outlines the key ingredients in the DWIA model.
Section~\ref{res} reports our calculations under two different kinematic 
arrangements.
Section~\ref{con} contains our concluding remarks.

\section{The Elementary Photoproduction Amplitude} \label{sec:elementary}

While dynamical models involving various
approximations for the Bethe-Salpeter equation
are becoming increasingly successful in the description 
of pion photoproduction, 
 the hadronic final state interaction in kaon photoproduction
 has usually been
left out \cite{ben:saghai,ben:williams,ben:mart95,ben:mart_s96}.
Neglecting the final meson-baryon interaction in the full meson
photoproduction $T$-matrix automatically leads to violation of 
unitarity since flux
going into inelastic channels has not been properly
accounted for.  Enforcing unitarity dynamically requires
solving a system of coupled channels with all possible final 
states.  In the case of $K^+ \Lambda$ photoproduction,
this arduous task has recently been accomplished
by Feuster and Mosel~\cite{feuster98,feuster99} using a K-matrix approach. 
However, such an amplitude is rather cumbersome to use
in reactions on nuclei. For our purpose, we 
therefore follow older models\cite{ben:saghai,ben:williams,ben:mart95,ben:mart_s96}
and choose an isobaric model without final-state interactions
which provides a simple tool to parameterize meson photoproduction 
off the nucleon.
Without rescattering contributions the $T$-matrix is simply
approximated by the driving term alone which is assumed to be given by 
a series of tree-level diagrams. The selected Feynman diagrams for the 
$s$-, $u$-, and $t$-channel contain some unknown coupling parameters to 
be adjusted in order to reproduce experimental data. Final state 
interaction is effectively absorbed in these coupling constants which 
then cannot easily be compared to couplings from other reactions.

One of the most contentious issues in the phenomenological 
description of kaon
photoproduction on the nucleon has been the choice of baryon resonances
in the production 
amplitude~\cite{ben:saghai,ben:williams,ben:mart95,feuster99,han99}.
Many authors have selected resonances that contribute
to the kaon production process by their relative contribution
to the overall $\chi^2$ of the fit~\cite{ben:saghai,ben:williams,han99}.
Our approach here is different: we wish to construct an amplitude
with a "minimal" number of resonances that is easy to handle in the
nuclear context.  We use the results of recent multichannel 
analyses~\cite{feuster98,feuster99,dytman2000,manley92}
as a guide to inform us of the most important resonances that decay into
$K \Lambda$ and $K \Sigma$ final states with a significant branching ratio.
Thus, in contrast to Refs.~\cite{ben:saghai,han99}, we do not include
spin 5/2 states in our amplitude since neither coupled-channels results
nor older partial-wave analyses\cite{saxon80,bell83} find their contributions
to be important.  Ultimately, only a multipole analysis will be able to 
clearly identify the resonances participating in kaon photoproduction.
Using input from the multichannel calculations by 
Refs.~\cite{feuster98,feuster99,dytman2000,manley92} we include the three 
resonances that have been found to 
decay noticeably into the $K \Lambda$ channel: $S_{11}$(1650), $P_{11}$(1710), 
and $P_{13}(1720)$. For $K \Sigma$ production we also allow contributions 
from the $S_{31}$(1900) and $P_{31}$(1910) resonances. 
Furthermore, we include not only the usual $1^-$ vector meson $K^*(892)$,
but also the $1^+$ pseudovector meson $K_1(1270)$ in the $t$-channel
since a number of studies~\cite{ben:saghai,ben:williams,ben:adelseck}
have found this resonance to give a significant contribution.

\subsection{Isospin Symmetry and Resonance Terms}
Following Refs.\cite{dennery,thom,deo}
we write the transition matrix of the reaction
\begin{eqnarray}
\gamma (p_\gamma) + N (p_N) \longrightarrow K(p_K) + Y(p_Y)
\end{eqnarray}
which stands for the following six reaction channels
\begin{eqnarray}
\gamma + p & \rightarrow & K^+ + \Lambda,  \label{react1} \\
\gamma + p & \rightarrow & K^+ + \Sigma^0, \label{react2} \\
\gamma + p & \rightarrow & K^0 + \Sigma^+, \label{react3} \\
\gamma + n & \rightarrow & K^0 + \Lambda,  \label{react4} \\
\gamma + n & \rightarrow & K^+ + \Sigma^-, \label{react5} \\
\gamma + n & \rightarrow & K^0 + \Sigma^0. \label{react6}
\end{eqnarray}
in the form of
\begin{eqnarray}
M_{\mathrm fi} &=& {\bar u}({\mbox{\boldmath ${p}$}}_Y,s_Y) 
  \sum_{i=1}^{4} A_i~M_i ~u({\mbox{\boldmath ${p}$}}_N,s_N) ~,
\label{eq:mfi}
\end{eqnarray}
where the Lorentz invariant matrices $M_i$ are given by
\begin{eqnarray}
 M_1 &=& \gamma_5 ~\epsilon\!\!/ p_\gamma\!\!\!\!\!/ ~~,\\
 M_2 &=& 2\gamma_5 (p_K\cdot\epsilon ~p_N\cdot p_\gamma - p_K\cdot p_\gamma~
         p_N\cdot\epsilon) ~,\\
 M_3 &=& \gamma_5 (p_K\cdot p_\gamma~ \epsilon\!\!/ - p_K\cdot \epsilon~
         p_\gamma\!\!\!\!\!/ ~ ) ~, \\
 M_4 &=& i\epsilon_{\mu\nu\rho\sigma}\gamma^\mu p_K^\nu \epsilon^\rho
         p_\gamma^\sigma ~.
\end{eqnarray}
The amplitudes $A_i$ are obtained from the Feynman diagrams of
Fig.~\ref{fig:diagram} by using the vertex factors and the propagators
given in Ref.~\cite{abw85,adelseck90}.  Casting the elementary operator
in the above form is convenient since it assures gauge invariance
even in the case of bound nucleons that the amplitude operates on
inside the nucleus in the framework of the impulse approximation.

To relate the hadronic coupling constants among the various isospin
channels we use isospin symmetry
\begin{eqnarray}
 g_{K^{+} \Lambda p} &=& g_{K^{0} \Lambda n} ~,\\
 g_{K^{+} \Sigma^{0} p} &=& -g_{K^{0} \Sigma^{0} n}
 = g_{K^{0} \Sigma^{+} p}/ \sqrt{2} = g_{K^{+} \Sigma^{-} n}/ \sqrt{2} ~,\\
 g_{K^{+} \Sigma^{0} \Delta^{+}} &=& g_{K^{0} \Sigma^{0}\Delta^{0}} 
 ~=~ -\sqrt{2} g_{K^{0} \Sigma^{+} \Delta^{+}} ~=~ 
 \sqrt{2} g_{K^{+} \Sigma^{-} \Delta^{0}} ~.
\end{eqnarray}
The electromagnetic couplings of the resonances to the proton and the neutron 
can be related by means of helicity amplitudes.
Following Ref.~\cite{feuster99} we can write the helicity amplitude 
of spin 1/2 resonances in terms of their coupling constants as
\begin{eqnarray}
 A_{1/2}^{(\pm)} &=& \mp \frac{1}{2m_N}
         \left(\frac{m_{N^*}^2-m_N^2}{2m_N}\right)^{1/2}~ eg_{N^*N\gamma} ~,
\end{eqnarray}
where the sign refers to the resonance parity of the resonance.
Therefore, the relation between spin 1/2 coupling constants for the production
on the proton and on the neutron is given by
\begin{eqnarray}
\frac{g_{N^{*0}n\gamma}}{g_{N^{*+}p\gamma}} &=&
\frac{A_{1/2}^n}{A_{1/2}^p} ~,
\end{eqnarray}

The Lagrangian for spin 3/2 resonances is, however, not unique. Using 
vertex functions as given in Ref.~\cite{abw85} we obtain the 
following relationships
\begin{eqnarray}
A_{1/2}^{(\pm)} &=& \frac{1}{2} \left[\frac{m_{N^*}\mp m_N}{{3m_N}(m_{N^*}
            \pm m_N)}\right]^{1/2} \left[ \frac{m_{N}}{m_{N^*}}
            eg^{(1)}_{N^*N\gamma} \pm \frac{1}{2} \left(\frac{m_{N^*}
            \mp m_N}{m_{N^*}
            \pm m_N}\right) eg^{(2)}_{N^*N\gamma}  \right]~,\\
A_{3/2}^{(\pm)} &=& \frac{1}{2} \left[\frac{m_{N^*}\mp m_N}{{m_N}(m_{N^*}
            \pm m_N)}\right]^{1/2} \left[ eg^{(1)}_{N^*N\gamma} -
            \frac{1}{2} \left(\frac{m_{N^*}\mp m_N}{m_{N^*}
            \pm m_N}\right)eg^{(2)}_{N^*N\gamma} \right]~.
\end{eqnarray}
and
\begin{eqnarray}
\frac{g^{(1)}_{N^{*0}n\gamma}}{g^{(1)}_{N^{*+}p\gamma}} &=&
\frac{\sqrt{3}A_{1/2}^n \pm A_{3/2}^n }{\sqrt{3}A_{1/2}^p \pm A_{3/2}^p } ~,\\
\frac{g^{(2)}_{N^{*0}n\gamma}}{g^{(2)}_{N^{*+}p\gamma}} &=&
\frac{\sqrt{3}A_{1/2}^n - ({m_N}/{m_{N^*}}) A_{3/2}^n }{\sqrt{3}A_{1/2}^p 
 - ({m_N}/{m_{N^*}}) A_{3/2}^p } ~
\end{eqnarray}
The numerical values for the $S_{11}(1650)$, $P_{11}(1710)$, 
and $P_{13}(1720)$ resonances are given in Table~\ref{tab:helicity}.

In $K^0$ photoproduction the transition moment $g_{K^{*+}K^+\gamma}$,
used in $K^+$ photoproduction, must be replaced by the neutral transition 
moment $g_{K^{*0}K^0\gamma}$. For both vector mesons, the $K^*$ and the $K_1$,
the transition moment is
related to the decay width by \cite{thom} 
\begin{eqnarray}
\Gamma_{K^*\rightarrow K\gamma}&=&\frac{1}{24}~\frac{|g_{K^*K\gamma}|^2}{4\pi 
M^2} \left[ m_{K^*}\left( 1-\frac{m_K^2}{m_{K^*}^2} \right) \right]^3 ~,
\label{eq:kdecay}
\end{eqnarray}
where $K^*$ refers to $K^*(892)$ or $K_1(1270)$, and $M=1$ GeV is used to
 make $g_{K^*K\gamma}$ dimensionless. 

The decay widths for $K^*(892)$ are well known, i.e.
\begin{eqnarray}
\Gamma_{K^{*+}\rightarrow K^+\gamma} &=& 50\pm 5 ~{\mathrm keV} ~,\\
\Gamma_{K^{*0}\rightarrow K^0\gamma} &=& 117\pm 10 ~{\mathrm keV} ~.
\end{eqnarray}
Thus, the transition moments are related by
\begin{eqnarray}
\label{eq:kmoment}
g_{K^{*0}K^0\gamma} &=& -1.53~ g_{K^{*+}K^+\gamma} ~,
\end{eqnarray}
where we have used the quark model prediction of Singer and Miller 
\cite{singer} in order to constrain the relative sign.

The decay widths of $K_1(1270)$ are, however, not well known. Nevertheless,
the ratio of the charged and neutral moment of  $K_1(1270)$ 
can be taken as a free parameter that is fixed by
the available data in the $p(\gamma,K^0)\Sigma^+$ channel.

In order to approximately account for unitarity corrections
at tree-level we include energy-dependent widths in the resonance propagators 
\begin{eqnarray}
\Gamma ({\mbox{\boldmath ${q}$}}) &=& \Gamma_{N^*}~
\frac{\sqrt{s}}{m_{N^*}}~
  \sum_{i} x_i \left( 
\frac{|{\mbox{\boldmath ${q}$}}_i|}{|{\mbox{\boldmath ${q}$}}_i^{N^*}|}
  \right)^{2l+1} \frac{D_l(|
  {\mbox{\boldmath ${q}$}}_i|)}{D_l(|{\mbox{\boldmath ${q}$}}_i^{N^*}|)} ~,
\label{eq:decay_width}
\end{eqnarray}
where the sum runs over the possible decay channels into a meson and 
a baryon with mass $m_i$ and $m_b$, respectively, and relative orbital
angular momentum $l$. In Eq.~(\ref{eq:decay_width}) 
$\Gamma_{N^*}$ represents the total decay width and 
$x_i$ is the relative branching ratio of the resonance into the $i$th channel.
The final state momenta are given by  
\begin{eqnarray}
|{\mbox{\boldmath ${q}$}}_i^{N^*}| &=&
  \left[\frac{(m_{N^*}^2-m_b^2+m_i^2)^2}{4m_{N^*}^2} -m_i^2\right]^{1/2} ~,
\end{eqnarray}
and
\begin{eqnarray}
|{\mbox{\boldmath ${q}$}}_i| &=& \left[\frac{(s-m_b^2+m_i^2)^2}{4s}
  -m_i^2\right]^{1/2} ~, 
\end{eqnarray}
while for the fission barrier factor 
$D_l({\mbox{\boldmath ${q}$}})$ we use the quark model result of 
Ref.~\cite{zhenping}
\begin{eqnarray}
D_l({\mbox{\boldmath ${q}$}}) &=& {\mathrm exp} \left( -\frac{
{\mbox{\boldmath ${q}$}}^2}{3\alpha^2} \right)~ ,
\end{eqnarray}
with $\alpha = 410 $ MeV.
The branching ratios, listed in Table~\ref{tab:branching}, are quite uncertain
for some of the partial decays. For this calculation we have used the ones
from Ref.~\cite{feuster98}. In general, we found our results to be fairly 
insensitive to this input.

%=============================================================
%                                       12/15/99--23:00
%===========HH: Begin ========================================
%
\subsection{Hadronic Form Factors and Gauge Invariance}

\begin{figure}[htbp]
  \begin{center}
    \leavevmode
    \psfig{figure=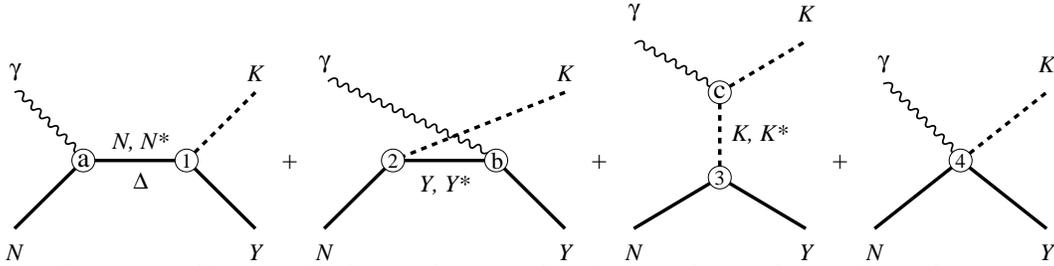,width=140mm}
    \caption{Feynman diagram for kaon photoproduction on the nucleon.
             Contributions from the $\Delta$ are only possible in $\Sigma$
             production. Electromagnetic vertices are denoted by (a), (b),
             and (c), hadronic vertices by (1), (2), and (3). The contact 
             diagram (4) is required in both PS and PV couplings in order 
             to restore gauge invariance after introducing hadronic form 
             factors. The Born terms contain the $N$, $Y$, $K$ intermediate 
             states and the contact term.\label{fig:diagram}}
  \end{center}
\end{figure}

It is a well-known fact that the sum of the first three photoproduction
diagrams---i.e., the sum of the $s$-, $u$-, and $t$-channel
diagrams---in Fig.~\ref{fig:diagram} is gauge-invariant only for bare
hadronic vertices with pure pseudoscalar coupling. Thus, in this most basic
case, the addition of a fourth contact-type graph in Fig.~\ref{fig:diagram}
is not necessary for preserving gauge invariance. In all other instances,
however, one needs additional currents to ensure gauge invariance
and thus current conservation. For bare hadronic vertices with pseudovector
coupling, this extra current is the well-known Kroll-Ruderman contact term \cite{kroll54}.

Irrespective of the coupling type, however, most isobaric models with bare vertices
show a divergence at higher energies, which clearly points to the need for introducing 
hadronic form factors to cut off this undesirable behavior. Recent calculations 
\cite{ben:saghai,ben:mart95} demonstrated that many models which are able to describe 
$(\gamma, K^+)$ experimental data tend to unrealistically overpredict
the $(\gamma, K^0)$ channel. The use of point-like particles disregards
the composite nature of nucleons and mesons, thus losing the full
complexity of a strongly interacting hadronic system.

To provide the desired higher-energy fall-off and still preserve the gauge invariance of the bare tree graphs,
the model of Ref.~\cite{ben:mart_s96} introduced a cut-off function by multiplying the entire photoproduction
 amplitude [see Eq.~(\ref{eq:mfi})] with an overall function of monopole form,
\begin{eqnarray}
F(\Lambda,t) &=& \frac{\Lambda^2-m_{K}^2}{\Lambda^2-t} ~,
\end{eqnarray}
where the cut-off mass $\Lambda$ was treated as a free parameter.   In spite of
successfully minimizing the $\chi^2$ while maintaining gauge invariance,
there is no microscopic basis for this approach since one cannot derive such an overall
factor from a field theory. 

Field theory clearly mandates that a correct description of vertex dressing effects must be 
done in terms of individual hadronic form factors for each of the three kinematic
situations given by the $s$-, $u$-, and $t$-channel diagrams of Fig.~\ref{fig:diagram}.
In a complete implementation of a field theory, the gauge invariance of the total
amplitude is ensured by the self-consistency of these dressing effects, by additional interaction
currents and by the effects of hadronic scattering processes in the final state \cite{haberzettl97}.
Schematically, the interaction currents and the final-state contributions can always be
written in the form of the fourth diagram of Fig.~\ref{fig:diagram}. In other words, 
the diagrammatic description of the photoproduction process given by this figure
is meaningful whether the vertices are bare or fully dressed; only the interpretation of the
individual diagrams changes: For bare particles, the diagrams correspond to the tree-level
bare Born terms only, whereas for fully dressed particles, the diagrams represent 
the topological structure of the {\it full} amplitude, with the first three graphs
depicting the fully dressed Born terms.

If one now seeks to describe the dressing of vertices on a more accessible, 
somewhat less rigorous, level, one introduces {\it phenomenological\/} form factors for 
the individual $s$-, $u$-, and $t$-channel vertices. Then, to ensure gauge invariance and to
remain close to the topological structure of the full underlying field theory, the simplest option is
to add contact-type currents which mock up the effects of the interaction currents and final-state 
scattering processes otherwise subsumed within the fourth diagram of Fig.~\ref{fig:diagram}.

One method to handle the inclusion of such phenomenological form factors has
been proposed by Ohta \cite{ben:ohta}. By making use of minimal substitution
Ohta has derived an additional current corresponding to the contact term of
Fig.~\ref{fig:diagram}. However, while Ohta's method does indeed restore gauge invariance,
its effect on the amplitude is the removal of any vertex dressing from the dominant electric 
contributions which---at least partially---undoes some of the
desirable effects of why dressed vertices needed to be introduced in the first place  \cite{hbmf98}.

Haberzettl has shown \cite{haberzettl97,hbmf98} that Ohta's method is too restrictive and that one
may retain the dressing effects suppressed by Ohta's approach by making use of the
fact that the longitudinal pieces of the gauge-invariance-preserving additional currents are 
only determined up to an arbitrary function $\tilde{F}$. (Of course, 
transverse currents are completely undetermined and arbitrary pieces 
can always be added with impunity.)
For practical purposes, one of the simplest choices \cite{haberzettl97,hbmf98} 
for this arbitrary function $\tilde{F}$ seems to be 
a linear combination of the form factors for the three kinematic situations in which
the dressed vertices appear , i.e., 
\begin{eqnarray} 
{\tilde F} &=& a_sF(\Lambda ,s) + a_uF(\Lambda ,u) + 
  a_tF(\Lambda ,t) ,\nonumber\\
  & &  ~~~ {\rm with~~} a_s+a_u+a_t= 1 ~,
  \label{habb_ff}
\end{eqnarray}
which introduces two more free parameters to be determined by fits to the experimental data. 
This choice has proven to be flexible and adequate for a 
good phenomenological description of experimental
data, and it is the choice adopted in the present work. In general, the results available
so far indicate that Haberzettl's method produces superior results compared to Ohta's approach
and has been used in all modern studies on kaon photoproduction~\cite{feuster99,han99,hbmf98} in 
an effective Lagrangian framework.

The inclusion of phenomenological form factors in the hadronic vertices of the Born terms in 
Fig.~\ref{fig:diagram} then leads to a modification of the four Born contributions $A_{i}^{\rm Born}$
that enter the respective coefficients $A_i$ of the photoproduction amplitude of Eq.~(\ref{eq:mfi}).
The additional contributions for each resonance are separately gauge invariant, by construction.
Following Refs.~\cite{haberzettl97,hbmf98},
the Born amplitudes for kaon photoproduction are given by
\begin{eqnarray}
A_{1}^{\rm Born} & = & -\frac{e g_{K Y N}}{s - m_{N}^{2}} \left(
Q_{N} +
\kappa_{N} \frac{m_{N} - m_{Y}}{2 m_{N}} \right) F(\Lambda,s) 
 - \frac{e g_{K Y N}}{u - m_{Y}^{2}} \left( Q_{Y} +
\kappa_{Y} \frac{m_{Y} - m_{N}}{2 m_{Y}} 
\right) F(\Lambda,u) \nonumber\\
&& - \left( 1 - \vert Q_{Y} \vert \right) \frac{e G_{K Y' N}}{u
- m_{Y'}^{2}}~ \frac{m_{Y'} - m_{N}}{m_{Y'} + m_{Y}}~ F(\Lambda,u) ~, \\
A_{2}^{\rm Born} & = & \frac{2e g_{K Y N}}{t - m_{K}^{2}} 
\left(\frac{Q_{N}}{s - m_{N}^{2}} + \frac{Q_{Y}}{u - m_{Y}^{2}} \right)
{\tilde F}~ ,\label{eq:ftilde} \\
A_{3}^{\rm Born} & = & \frac{e g_{K Y N}}{s - m_{N}^{2}}~
\frac{\kappa_{N} F(\Lambda,s)}{2 m_{N}} - \frac{e g_{K Y N}}{u - 
m_{Y}^{2}}~ \frac{\kappa_{Y} F(\Lambda,u)}{2 m_{Y}} 
 - (1 - \vert Q_{Y} \vert) \frac{e G_{K Y' N}}{u - m_{Y'}^{2}}~ 
 \frac{ F(\Lambda,u)}{m_{Y'} + m_{Y}}~ , \\
A_{4}^{\rm Born} & = & \frac{e g_{K Y N}}{s - m_{N}^{2}}~
\frac{\kappa_{N} F(\Lambda,s)}{2 m_{N}} +
\frac{e g_{K Y N}}{u - m_{Y}^{2}}~ \frac{\kappa_{Y} F(\Lambda,u)}{2 m_{Y}}
+ (1 - \vert Q_{Y} \vert)\frac{e G_{K Y' N}}{u - m_{Y'}^{2}}~
\frac{F(\Lambda,u)}{m_{Y'} + m_{Y}}~ ,
\end{eqnarray}
where $Q_{N}$ and $Q_{Y}$ denote the
charge of the nucleon and the hyperon in $+e$ unit, while $\kappa_N$,
$\kappa_Y$, and $\kappa_T$ indicate the anomalous magnetic moments of the
nucleon, hyperon, and the transition of $\Sigma^0 \Lambda$. It is
understood that $Y'=\Sigma^0 ~[\Lambda]$ for $K\Lambda ~[K\Sigma^0]$
production. 
As can be seen here, the function $\tilde{F}$ governs the fall-off behavior of the $A_{2}^{\rm Born}$
term which describes the dominant electric contributions of the Born terms. (Note here that Ohta's choice
corresponds to $\tilde{F}=1$ \cite{hbmf98} and thus provides no cut-off for higher energies for this term.)

Finally, we mention that for practical purposes we have introduced a slightly different notation
for the linear combination in Eq.~(\ref{habb_ff}), namely
\begin{eqnarray}
{\tilde F} &=& \sin^2\Theta_{\rm hd} \cos^2\Phi_{\rm hd} F(\Lambda,s) +
        \sin^2\Theta_{\rm hd}\sin^2\Phi_{\rm hd}F(\Lambda,u) + 
        \cos^2\Theta_{\rm hd} F(\Lambda,t) ~,
\label{eq:fhabtilde}
\end{eqnarray}
where the combination of trigonometric functions ensures the correct
normalization of $\tilde{F}$. Both $\Theta_{\rm hd}$ and 
$\Phi_{\rm hd}$ are obtained from the fit and quoted in Table \ref{tab:cc}.
For the functional dependence of the form factor we use a
covariant vertex parameterization without singularities 
on the real axis,
\begin{eqnarray}
F(\Lambda ,q^2) &=& \frac{\Lambda^4}{\Lambda^4+\left( q^2-m^2 \right)^2} ~,
\end{eqnarray} 
with $q^2=s$, $t$, or $u$, and $m$ being the mass of the intermediate
particle of the respective diagram.

%
%===========HH: End ==========================================
%=============================================================

\subsection{Comparison to Photoproduction Data on the Nucleon}

We have performed a combined fit to all differential
cross section and  recoil polarization data of 
$p(\gamma,K^+)\Lambda$ and $p(\gamma,K^+)\Sigma^0$.
The present data base includes the new {\footnotesize SAPHIR} 
data set up to $W=2.1$ GeV \cite{saphir98}, but excludes the 
older {\footnotesize SAPHIR} data, published
in Ref.~\cite{bockhorst}, which have significantly
larger error-bars. 
Both statistical and systematic errors are included;
for the small number of old data that did not report systematic errors, we
added a 10\% uncertainty 
to their error-bars.  With the upcoming high-precision Jefferson Lab results
the data base is about to experience further significant improvements. 
The $p(\gamma,K^0)\Sigma^+$ channel is included later, since data for this 
channel have large error bars, and therefore do not strongly influence
the fit. 
%Furthermore, the available $K^0 \Sigma^+$ data reported in
%Ref.~\cite{bennhold97} were preliminary; the final data have been reanalyzed
%and should appear soon\cite{goers99}.
 
The results of our fits are summarized in Table \ref{tab:branching}.
We compare our present study to an older model \cite{ben:mart_s96}
which employed an overall hadronic form factor and did not contain
the $P_{13}$(1720) and the $K_1$(1270) states. The significant improvement
in $\chi^2$ comes mostly from including the $P_{13}$(1720) in
the $K \Lambda$ channel.  A further reduction in $\chi^2$
results from allowing the non-resonant background terms to have a different
form factor cut-off than the $s$-channel resonances.  For the former, the 
fit produced a soft value of about 800 MeV, leading to a strong suppression of the 
background terms while the resonance cut-off is determined to be 1.89 GeV.
This combination leads to a reaction mechanism
 which is resonance dominated in all isospin channels.
Table \ref{tab:branching} reveals that the coupling
ratio $K_1^0K^0\gamma / K_1^+K^+\gamma$ is obtained with 
large uncertainty. This comes as no surprise since the data
in the $p(\gamma,K^0)\Sigma^+$ channel have large error-bars;
we predict the ratio of the decay widths to be
\begin{eqnarray}
\frac{\Gamma_{K_1^0\rightarrow K^0\gamma}}{\Gamma_{K_1^+\rightarrow
K^+\gamma}} &=& 0.068 \pm 0.110 ~.
\end{eqnarray}

\begin{figure}[!ht]
  \begin{center}
    \leavevmode
    \psfig{figure=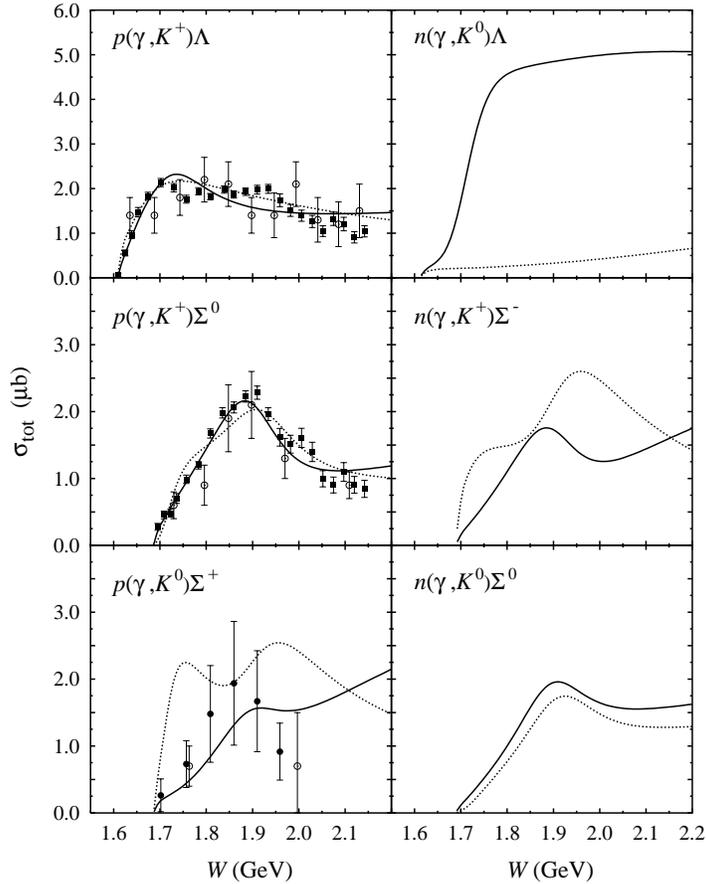,width=100mm}
    \caption{\label{fig:total} Total cross sections
        for the six isospin channels of kaon photoproduction on the 
        nucleon calculated at tree level. The solid curve shows Set II 
        of Table \protect\ref{tab:cc} 
        while the dotted line shows the older model, Set I 
        of Table \protect\ref{tab:cc}. The new {\scriptsize SAPHIR} 
        data \protect\cite{saphir98} are denoted by the solid 
        squares, old data \protect\cite{old_data} are shown by the 
        open circles. Solid circles are the data for $K^0\Sigma^+$ production 
        from Ref. \protect\cite{benn97}.}
  \end{center}
\end{figure}

Fig.~\ref{fig:total} compares total cross section data for the 
three different
$K^+$ photoproduction reactions on the proton. For $p(\gamma,K^+)\Lambda$
one can see a possible signal for a cusp effect around $W = 1710$ MeV, 
indicating
the opening of the $K \Sigma$ channel. The steep rise of the $K^+ \Lambda$
data at threshold is indicative of a strong $s$-wave. 
The $K^+ \Lambda$ data
reveal an interesting structure around $W=1900$ MeV.  Our model fits currently
do not reproduce this feature since there is no well-established (3- and 
4-star) $I=1/2$ state at this energy. 
However, Ref. \cite{capstick98} predicts a missing $D_{13}$ at 1960 MeV that 
has a large branching ratio both into the $\gamma N$ and  the $K \Lambda$ 
channel. In order to study this structure more closely, Ref.~\cite{mart99} 
has included a $D_{13}$ resonance but allowed the mass and the width of
the state to vary as free parameters. A significant 
reduction in $\chi^2 / N$ for a mass of 1895 MeV and a total width of 372 MeV
was achieved. Because of its uncertain nature, 
this state is not included in the present calculation.

The $K^+ \Sigma^0$ data rise more slowly at threshold,
suggesting $p$- and $d$-wave, rather than $s$-wave, dominance.  
Furthermore, there is a clear evidence for
a resonance structure around $W = 1900$ MeV.
There is indeed a cluster of six or seven $\Delta$ resonances with spin
quantum numbers 1/2, 3/2, 5/2 and 7/2; it is at this energy
that the total $K \Sigma$ cross section reaches its maximum.
Disentangling these overlapping resonance contributions will 
require a multipole analysis.  The $K^0 \Sigma^+$ data have large error 
bars, thus few conclusions can be drawn at this time.  Nevertheless,
they appear to have a similar resonance structure around
1900 MeV. No data are available for production on the
neutron, this situation will be remedied by the ongoing analysis
of the g2 data at Jefferson Lab.

\begin{figure}[htbp]
  \begin{center}
    \leavevmode
    \psfig{figure=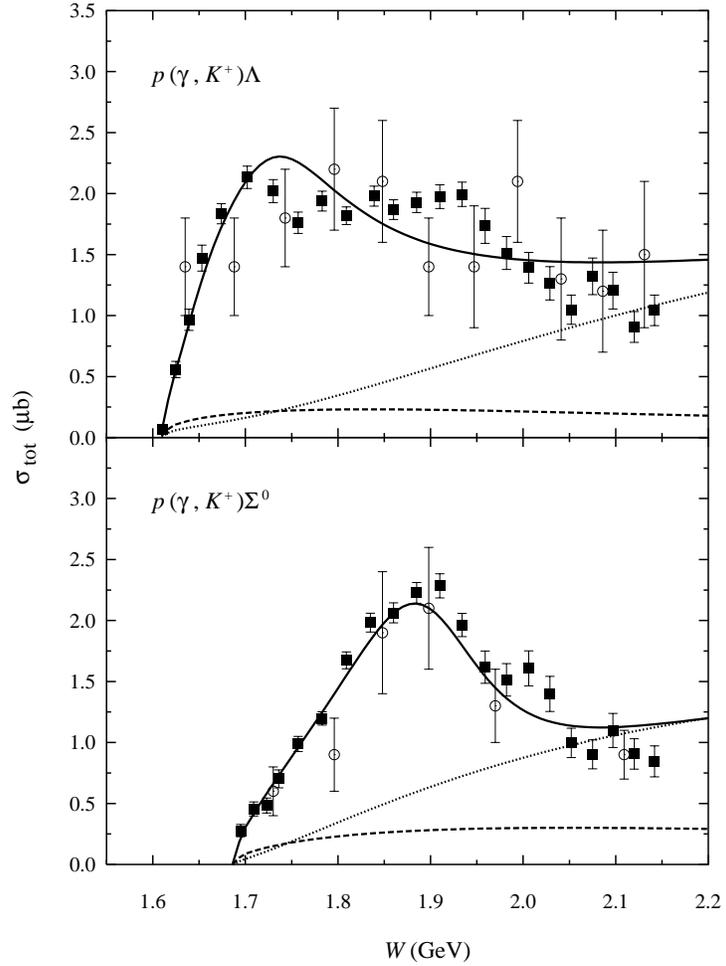,width=100mm}
    \caption{Contribution of the Born terms (dashed lines), 
             Born + $K^*$ + $K_1$ terms
             (dotted lines), and full operator (solid lines)
             to the total cross section of the 
             $p(\gamma ,K^+)\Lambda$ and $p(\gamma ,K^+)\Sigma^0$ channels.
             The notation of the data is as in Fig.~\ref{fig:total}.}
    \label{fig:contrib}
  \end{center}
\end{figure}

Fig.~\ref{fig:contrib} displays the dominance of the resonances
in the production process. Due to the presence of hadronic form factors
the Born terms contribute about 10-20\% 
to the total cross sections and do not exhibit the divergent behavior
well known from earlier 
studies~\cite{Rosenthal88,ben:saghai,ben:williams,ben:mart95,abw85,adelseck90}.
At higher energies, the vector meson
$t$-channel terms become large, indicating that in this 
energy regime corrections of the form found in Regge descriptions
\cite{guidal} may have to be applied. This also suggests that the range of
applicability of isobar models based on effective Lagrangians may be
limited to an energy up to $W = 2.2-2.5 GeV$; beyond this energy descriptions
based on Regge trajectories may become more appropriate.  This transition between
the s-channel resonance regime and the t-channel Regge region involves the concept
of duality and is currently subject of intense study\cite{carlson2000}.

\begin{figure}[htbp]
  \begin{center}
    \leavevmode
    \psfig{figure=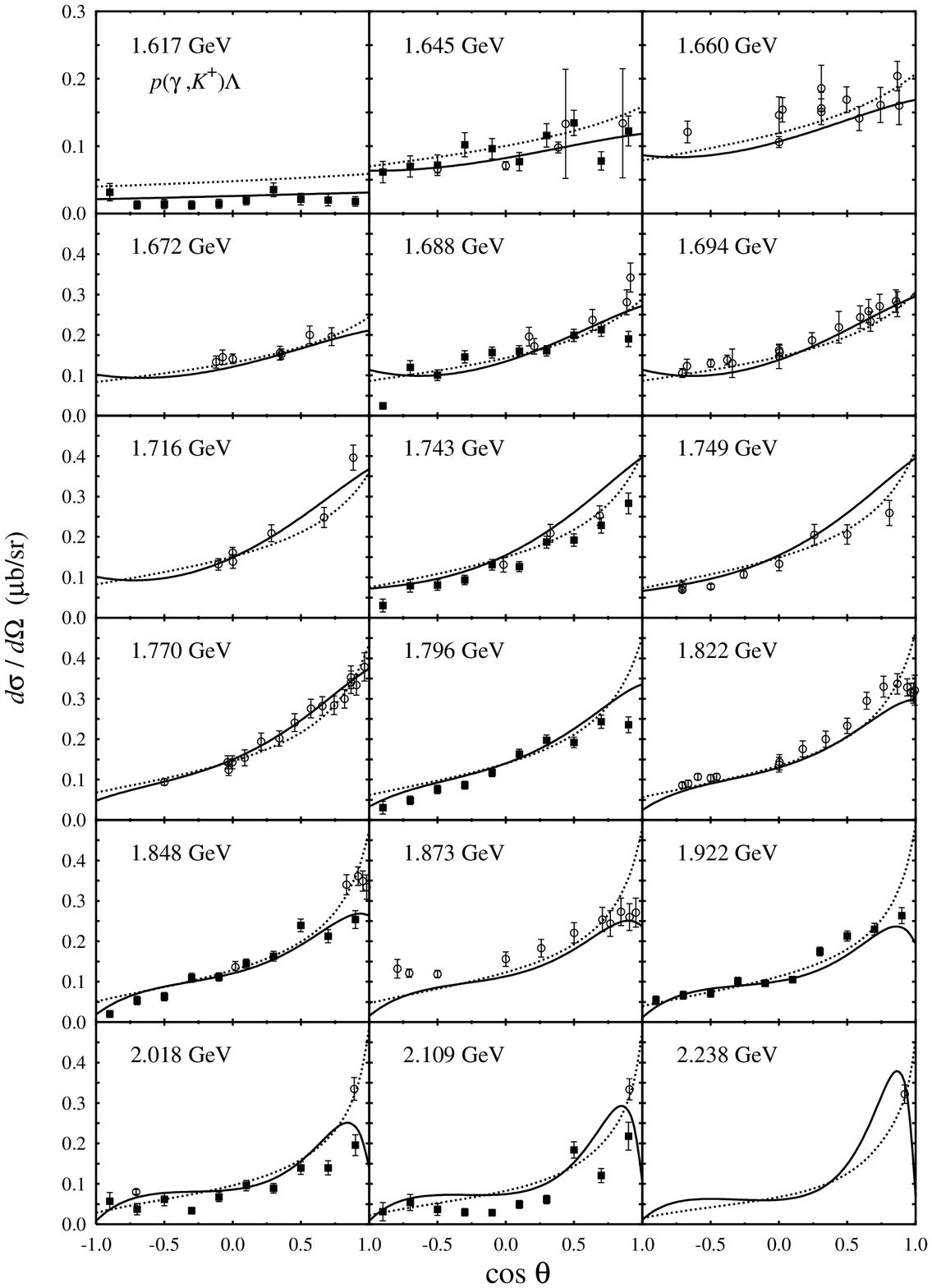,width=140mm}
    \caption{Differential cross section for $p(\gamma ,K^+)\Lambda$ channel.
             The notation of the curves is as in Fig.~\ref{fig:total}.
             The total c.m. energy $W$ is shown in every panel.}
    \label{fig:difkpl}
  \end{center}
\end{figure}

Fig.~\ref{fig:difkpl} shows the differential cross section of the
$p(\gamma,K^+)\Lambda$ channel for the two models listed in
Table \ref{tab:branching}. At threshold, the process is dominated by $s$-wave, 
due mostly to Born terms but also to the $S_{11}$(1650).
Around 1700 MeV we find the onset of a forward-backward asymmetry
due to p-waves coming from the $P_{11}$(1710) and $P_{13}$(1720) states.
At higher energies we find strong forward
peaking similar to the $p(\pi^-,K^0)\Lambda$ case
that can be attributed to the $K^*$ contribution\cite{feuster98}.
While the total cross section data were equally well
reproduced by both models, Set II is superior in 
describing the differential cross sections, especially
at threshold. It demonstrates that amplitudes using an overall
form factor of the form of Eq. (30) do not have enough kinematic flexibility
to accommodate the entire energy region under consideration. Similar results 
have been found for the gauge prescription according to 
Ohta\cite{han99,elba98}.

\begin{figure}[!ht]
  \begin{center}
    \leavevmode
    \psfig{figure=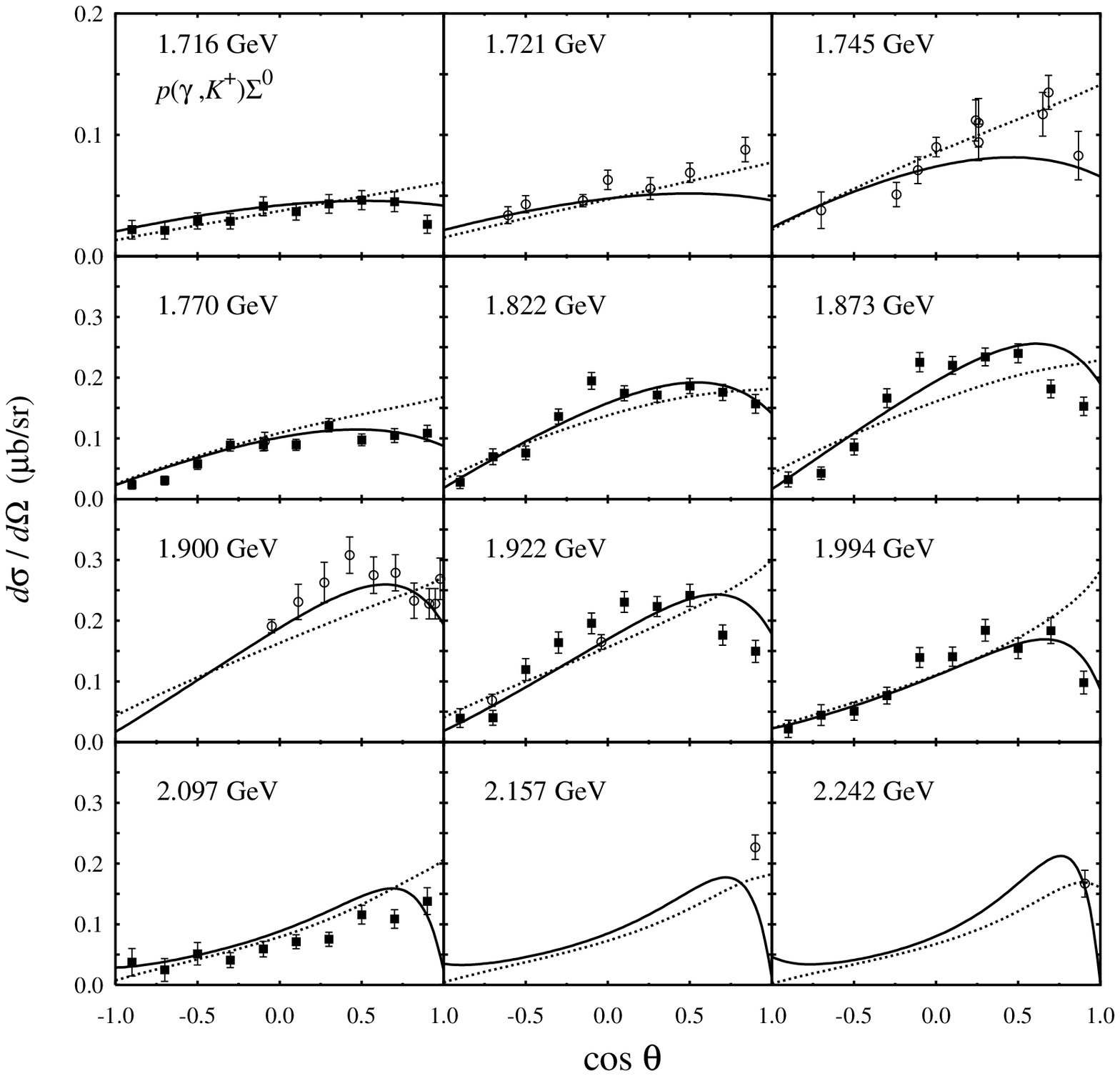,width=120mm}
    \caption{Differential cross section for 
             $p(\gamma ,K^+)\Sigma^0$ channel.
             The notation of the curves is as in Fig.~\ref{fig:total}.
             The total c.m. energy $W$ is shown in every panel.}
    \label{fig:difkps0}
  \end{center}
\end{figure}

The comparison of the two models with the $p(\gamma,
K^+)\Sigma^0$ data is shown in Fig.~\ref{fig:difkps0} 
from threshold up to 2.2 GeV. In contrast to $K^+ \Lambda$
photoproduction, this channel contains significant $p$- and $d$-wave 
contributions
already at threshold. This points to the $P_{11}$(1710) state as an important
resonance in low-energy $K \Sigma$ production; here the $S_{11}$(1650) 
lies below threshold. This finding is consistent with a recent 
study \cite{ben:sibirtsev98}
of $K \Sigma$ production in $NN$ scattering, $N N \rightarrow N K \Sigma$,
where the $P_{11}$(1710) state was identified as a major contribution.
Furthermore, recent coupled-channel analyses by Waluyo et al.\cite{waluyo2000}
identify the $P_{11}$(1710) state as the dominant resonance in low-energy
$\pi N \rightarrow K \Sigma$ reactions with a branching ratio of
$P_{11}(1710) \rightarrow K \Sigma$ of 32 MeV.  In contrast to $K \Lambda$
photoproduction the forward peaking is less pronounced, due in part
to smaller $g_{K^* N \Sigma}$ coupling constants. Around W = 1900 MeV, the cross section
is dominated by two isospin 3/2 states, the $S_{31}$(1900) and the $P_{31}$(1910).

\begin{figure}[!ht]
  \begin{center}
    \leavevmode
    \psfig{figure=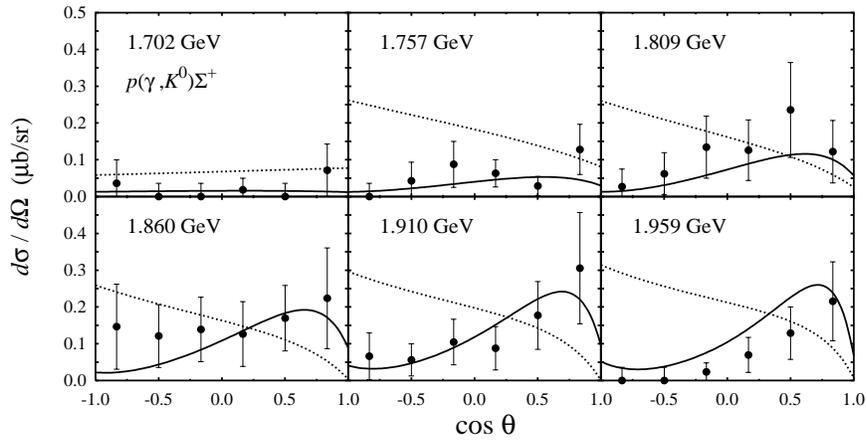,width=120mm}
    \caption{\label{fig:difk0sp} Differential cross section 
             for $p(\gamma ,K^0)\Sigma^+$ channel.
             Data are from Ref.~\protect\cite{benn97}.
             Notation is as in Fig. \ref{fig:total}.
             The total c.m. energy $W$ is shown in every panel.}
  \end{center}
\end{figure}

Fig.~\ref{fig:difk0sp} compares the two models for the
$p(\gamma,K^0)\Sigma^+$ channel. 
The dramatically different behavior between the two models
is due mostly to the different
gauge prescriptions used since this influences the relative contribution
of the background terms.
As mentioned above, Set I used an overall hadronic form factor that 
multiplied the entire amplitude, while Set II employs 
the mechanism by Haberzettl, which is preferred by the data.

\begin{figure}[!ht]
  \begin{center}
    \leavevmode
    \psfig{figure=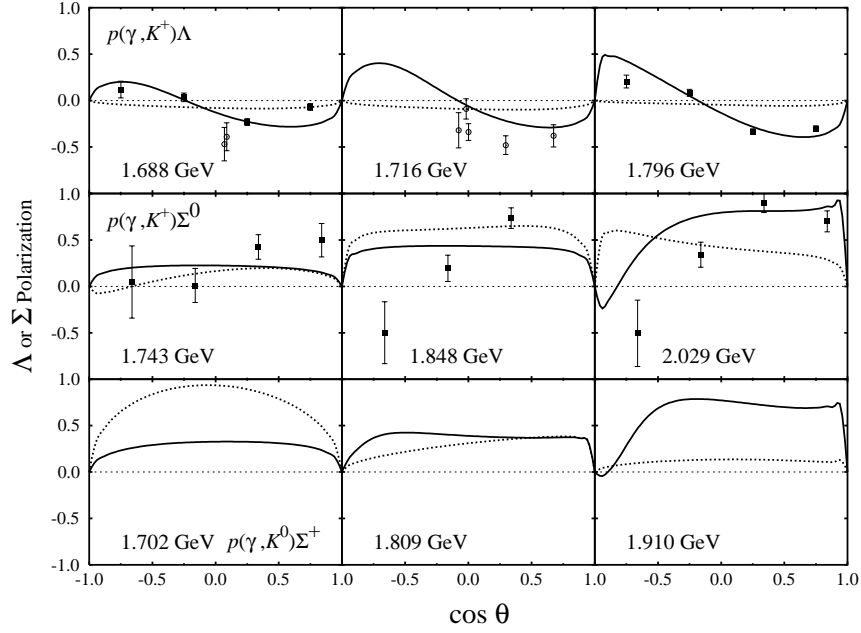,width=120mm}
    \caption{\label{fig:recoil} $\Lambda$ and $\Sigma$ recoil polarization 
          for $p(\gamma ,K^+){\vec Y}$. Notation is as in Fig. 
          \ref{fig:total}. The total c.m. energy $W$ is shown in every panel.}
  \end{center}
\end{figure}

The recoil polarization for the three reaction channels
on the proton is shown in Fig.~\ref{fig:recoil}.
For the $K^+ \Lambda$ data
we find good agreement using Set II of Table \ref{tab:branching}, while the older
model (Set I) gives almost zero polarization throughout this energy range.
We point out that the {\footnotesize SAPHIR} data are binned in 
large angular and energy intervals.
The main reason for this dramatic difference is the more prominent role
that the resonances play in the present model, defined by Set II.
In the case of $K^+ \Sigma^0$ photoproduction the models fails
to reproduce the polarization data.
Since the recoil polarization observable
is sensitive especially to the imaginary parts of the amplitudes this 
discrepancy suggests that we do not have the correct resonance input 
for the $K \Sigma$ channel.

\begin{figure}[!ht]
  \begin{center}
    \leavevmode
    \psfig{figure=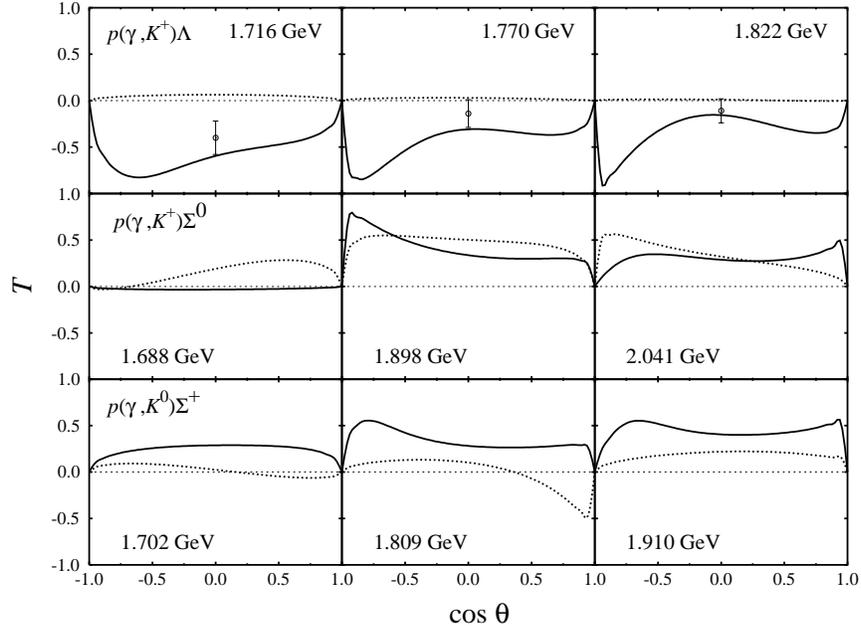,width=120mm}
    \caption{\label{fig:target} Target asymmetry for the reaction 
        ${\vec p}(\gamma ,K^+) Y$.
             Notation is as in Fig. \ref{fig:total}. 
             The total c.m. energy $W$ is shown in every panel.}
  \end{center}
\end{figure}

In Fig.~\ref{fig:target} we show the target asymmetry for the same three 
production processes at selected kinematics.  Only three data
points are available for $K^+ \Lambda$ production, which we did not include 
in the fit. At threshold the target asymmetry calculated with Set II
is predicted to be sizable for $K^+ \Lambda$ production
but small for the two $K \Sigma$ production channels.  Similar to 
the $\Lambda$ recoil polarization in Fig.~\ref{fig:recoil} Set I predicts
a zero asymmetry for $K^+ \Lambda$ production for the first 200 MeV above threshold.
At higher energies significant asymmetries are obtained for the $K \Sigma$
production reactions. However, the differences between Sets I and II are not
too large, suggesting that this may not be the most appropriate observable
to discriminate between the two models.

\begin{figure}[htbp]
  \begin{center}
    \leavevmode
    \psfig{figure=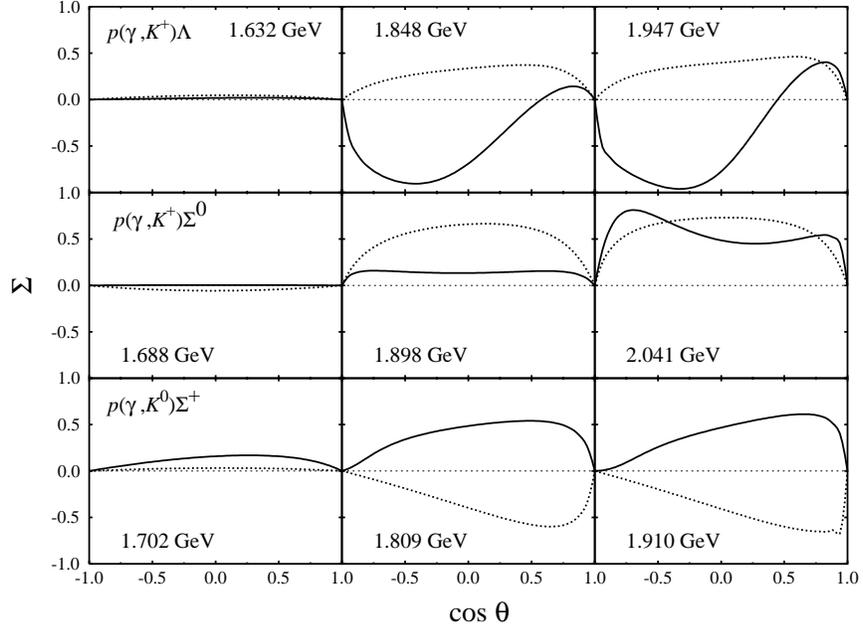,width=120mm}
    \caption{Photon asymmetry of $p({\vec \gamma},K^+)Y$.
             Notation is as in Fig. \ref{fig:total}.
             The total c.m. energy $W$ is shown in every panel.}
    \label{fig:polph}
  \end{center}
\end{figure}

The last figure in this section involves polarized photons.  
The beam asymmetry $\Sigma$ can be measured with linearly polarized
photons, which will become available at Jefferson Lab within a year. 
As shown in Fig. \ref{fig:polph} 
this asymmetry is almost zero near threshold for all three channels
but becomes sizable at higher energies. We find large differences between
the two models, suggesting that this is an ideal observable to distinguish 
between different dynamical inputs.  This observation was also made in
ref.\cite{mart99} where it was found that the polarized photon asymmetry is 
well suited to shed light on the nature of the "missing" $D_{13}$ resonance 
around W = 1900 MeV in $K^+ \Lambda$ production.

Concluding this section, we reemphasize the potential of polarization
observables to discriminate between models that use different dynamical inputs.
The primary dynamical ingredients in all effective Lagrangian descriptions of
kaon photoproduction are the nonresonant background terms and the s-channel 
resonances.  As the need for hadronic form factors at these energies has become
widely recognized a choice must be made with regard to the restoration of 
gauge invariance.  While the method by Haberzettl has a clear 
field-theoretical foundation it is desirable to establish its preference
phenomenologically as well.  As demonstrated in the above figures,
polarization observables play a crucial role. Once a proper 
description of the Born terms is accomplished the resonances can
be investigated in detail.
We point out that the use of polarized electron beams 
produces circularly polarized photons, which in combination with
the hyperon recoil polarization allows for the 
measurement\cite{reinhard} of
the beam-recoil double-polarization 
observables $C_x$ and $C_z$. Such data have already been taken and
are currently being analyzed\cite{reinhard}.  Furthermore,
the availability of linearly polarized photons at JLab will allow
the measurement\cite{sanabria2000} of the beam-recoil observables $O_x$, $O_z$ and 
$O_y$ (which is identical to -$T$, the polarized target asymmetry).
Such a set of observables constitutes an almost complete experiment
and should allow a multipole analysis that can aid in the determination
of the resonances and the extraction of resonance parameters.

\section{The DWIA framework}
\label{dwia}

Now we consider the kaon photoproduction process 
on a nuclear target in the DWIA model. 

\subsection{Differential Cross Section}
\label{diff}

Working in the laboratory frame where the target is at rest,    
we difine the coordinate system such that
the $z$-axis is along the photon direction ${\mbox{\boldmath ${p}$}}_\gamma$,
and the $y$-axis is along 
${\mbox{\boldmath ${p}$}}_\gamma \times {\mbox{\boldmath ${p}$}}_K$ with 
the azimuthal angle of the kaon chosen as $\phi_K = 0$.
The kinematics of the reaction are determined by
\begin{equation}
{\mbox{\boldmath ${p}$}}_\gamma={\mbox{\boldmath ${p}$}}_K +
{\mbox{\boldmath ${p}$}}_Y + {\mbox{\boldmath ${p}$}}_m ~, 
\label{mcon}
\end{equation}
\begin{equation}
E_\gamma +M_i = E_K +E_Y +M_f + T_m ~.
\label{econ}
\end{equation}
Here ${\mbox{\boldmath ${p}$}}_m$ is the missing momentum in the reaction and 
$T_m={p_m^2 / 2M_f}$ is the recoil kinetic energy.
The excitation energy of the residual nucleus is included in $M_f$.
The missing energy $E_m$ in the reaction is defined by 
$E_m=M_f-M_i+m_N=E_\gamma-E_K -E_Y-T_m+m_N$ where 
$m_N$ is the mass of the nucleon.
For real photons, $|{\mbox{\boldmath ${p}$}}_\gamma |=E_\gamma$.
In the impulse approximation, 
the reaction is assumed to take place on a single bound nucleon whose 
momentum and energy are given by 
${\mbox{\boldmath ${p}$}}_i =-{\mbox{\boldmath ${p}$}}_m$ 
and $E_i=E_K +E_Y -E_\gamma$.  
This seems the most sensible choice for the bound nucleon, since
all other particles are observed in the laboratory.
With such constraints on $E_i$ and ${\mbox{\boldmath ${p}$}}_i$, 
the struck nucleon is in general off its mass shell, except right
on top of the quasifree peak (${\mbox{\boldmath ${p}$}}_m=0$).  
Since we are mostly interested in the quasifree region, the
off-shell effects are expected to be small.

The reaction is {\em quasifree}, meaning that the magnitude of 
${\mbox{\boldmath ${p}$}}_m$ can have a wide range, including zero.
Since the reaction amplitude is proportional to the Fourier transform 
of the bound state single particle wavefunction, it falls off quickly 
as the momentum transfer ${\mbox{\boldmath ${p}$}}_m$ increases. 
Thus we will restrict ourselves to the low $p_m$ region ($<$ 500 MeV)  
where the nuclear recoil energy ($T_m$) can be safely neglected 
for nuclei of $A > 6$.

The differential cross section can be written as 
\begin{equation}
\frac{d^3 \sigma}{dE_K\,d\Omega_K\,d\Omega_Y}=
{C \over 2(2J_i+1)}\;
\sum_{\alpha,\lambda,m_s}\frac{S_\alpha}{2(2j+1)}
|T(\alpha,\lambda,m_s)|^2.
\label{coin}
\end{equation}
The kinematic factor is given by
\begin{equation}
C=
{ M_f m_Y\, |{\mbox{\boldmath ${p}$}}_K|\,|{\mbox{\boldmath ${p}$}}_Y| \over 
4(2\pi)^5 
|E_Y +M_f+T_m -E_Y \,{\mbox{\boldmath ${p}$}}_Y
\cdot({\mbox{\boldmath ${p}$}}_\gamma-{\mbox{\boldmath ${p}$}}_m)/p_Y^2| }.
\end{equation}
The single particle matrix element is given by
\begin{equation}
T(\alpha,\lambda,m_s) = \int d^3 r\,
\Psi^{(+)}_{m_s}({\mbox{\boldmath ${r}$}},-{\mbox{\boldmath ${p}$}}_Y)\;
\phi^{(+)}_K({\mbox{\boldmath ${r}$}},-{\mbox{\boldmath ${p}$}}_K)\;
t_{\gamma K}(\lambda, {\mbox{\boldmath ${p}$}}_\gamma ,
{\mbox{\boldmath ${p}$}}_i , {\mbox{\boldmath ${p}$}}_K ,
{\mbox{\boldmath ${p}$}}_Y )\;
\Psi_{\alpha}({\mbox{\boldmath ${r}$}})\; {\rm exp}
(i{\mbox{\boldmath ${p}$}}_\gamma\cdot {\mbox{\boldmath ${r}$}}).  
\label{3d}
\end{equation}
In the above equations, $J_i$ is the target spin, 
$\alpha=\{nljm\}$ represents the single particle states,
$S_\alpha$ is called the spectroscopic factor,
$\lambda$ is the photon polarization, $m_s$ is
the spin projection of the outgoing nucleon, $\Psi^{(+)}_{m_s}$
and $\phi^{(+)}_K$ are the distorted wavefunctions with outgoing
boundary conditions, $\Psi_{\alpha}$ is the bound nucleon wavefunction,
and $t_{\gamma K}$ is the kaon photoproduction operator, discussed 
in the previous section.

In addition to cross sections, we also compute polarization observables. 
One is the photon asymmetry defined by
\begin{equation}
A_\gamma=
{ 
{d^3 \sigma}_{\perp}-{d^3 \sigma}_{\parallel}
\over
{d^3 \sigma}_{\perp}+{d^3 \sigma}_{\parallel}
},
\end{equation}
where $\perp$ and $\parallel$ denote the perpendicular and parallel
photon polarizations relative to the production plane ($x$-$z$ plane).
Another is the hyperon recoil polarization (also called analyzing power) 
defined by 
\begin{equation}
A_Y=
{ 
{d^3 \sigma}_{\uparrow}-{d^3 \sigma}_{\downarrow}
\over
{d^3 \sigma}_{\uparrow}+{d^3 \sigma}_{\downarrow}
},
\end{equation}
where $\uparrow$ and $\downarrow$ denote the polarizations 
of the outgoing hyperon relative to the $y$-axis.
We have used the short-hand notation
$d^3 \sigma \equiv d^3 \sigma/dE_K\,d\Omega_Kd\Omega_Y$ with 
appropriate sums over spin labels implied. Note that $A_Y$ is obtained 
for free experimentally since the produced hyperon is self-analyzing, 
while the measurement of $A_\gamma$ requires  polarized photon beams.

\subsection{Nuclear Structure Input}
\label{nucin}
The dependence of the reaction on nuclear structure is minimal.
It enters through the spectroscopic factor $S_\alpha$
and the single particle bound wavefunction.
The former is an overall normalization factor whose value can be 
taken from electron scattering. It cancels out in polarization observables.
For the latter we use harmonic oscillator wavefunctions. For the sake of consistency
one should use bound-state wave functions originating from a similar potential well
as the outgoing hyperons. However, for the quasifree region we are interested in
the difference is negligible.

\subsection{Kaon-Nucleus Interaction}
\label{kopt}

Unlike the $\pi N$ interaction,
the $K^+N$ interaction is rather weak on the hadronic scale.
Because of strangeness conservation, there are no hyperon resonances 
in the $K^+N$ system, nor any inelastic channels with the obvious 
exception of $(K^+,K^0)$ charge exchange on the neutron.
The large medium effects due to $\pi NN\rightarrow NN$ annihilation and 
and $\Delta$ propagation in the $\pi$-nucleus system are  absent 
from the $K^+$-nucleus scattering.
Consequently, the low-energy $K^+N$ interaction can be understood by a 
simple background scattering with a smooth energy dependence.
To generate the distorted waves, we solved the Klein-Gordon equation 
with a first-order optical potential constructed from the 
elementary $K^+N$ amplitudes by a simple $t\rho$ 
approximation~\cite{Bennhold89}. 
For $K^0$, we used the same potential as for $K^+$ as a starting
point, since little is known about the $K^0$ nucleus
interaction.  In principle, such information can be obtained
by measuring kaon charge exchange on nuclei.  Better optical potentials,
such as the one developed in Ref.\cite{kamalov},
should be incorporated in future studies; however,
for the present purpose of an exploratory study
these potentials are sufficient.

\subsection{Hyperon-Nucleus Interaction}
\label{ypot}

Very few optical potentials have been constructed
to describe hyperon-nucleus scattering, mostly due to lack of data. 
Here, we employ the global optical model by Cooper 
{\it et al}~\cite{Cooper94}.
It was built upon a global nucleon-nucleus Dirac optical 
potential~\cite{Cooper93} that successfully describes the nucleon data over a
wide range of nuclei and energies.
It provides the strengths and shapes for the real and imaginary parts
of the nucleon-nucleus scalar and vector potentials.
Then, a number of assumptions were made to deduce the hyperon-nucleus 
optical potentials.
First, it was assumed that the real parts of the hyperon scalar and 
vector potentials scale down by  factors $\alpha_s$ and
$\alpha_v$ motivated by the constituent quark model, and that
the imaginary parts scale down like the square of the same factors.
Second, a tensor coupling term was included in the potential.
The coupling was again motivated by the constituent quark model:
$f=-g$ for the $\Lambda$ and $f=+g$ for the $\Sigma$.
Here $f$ is the strength of the tensor coupling of the hyperon to the
$\omega$ meson and $g$ is the corresponding vector coupling.
The tensor coupling term was neglected for the nucleon since 
the $\omega N$ coupling constant is small.
The inclusion of the tensor terms makes the $\Lambda N$ interaction  
approximately spin-independent as suggested by the $\Lambda$ 
hypernuclear data, 
and the $\Sigma N$ interaction maximally spin-dependent.
Third, for $\Sigma$, an additional contribution due to 
$\Sigma N \rightarrow \Lambda N$ conversion is known to affect the 
imaginary part of the potential, and it was parameterized by 
adding a certain amount, $\Delta V_s$, to the imaginary part of the 
scalar potential.
The soundness of these assumptions may deserve further study, 
they nonetheless provide a basis for this qualitative
study of the hyperon-nucleus interaction.

This model was applied to bound hypernuclear systems and was found to 
give a reasonable description of the experimental data~\cite{Mares94}. 
The parameters were then adjusted slightly to reproduce the  
data more quantitatively.  In the case of the $\Sigma$, the model was also constrained by the 
existing information from $\Sigma^-$ atoms and from $\Sigma N$ scattering.
In this study, we will use the following parameters for $\Lambda$:
\begin{equation}
\alpha_s=0.621, \;\;
\alpha_v=0.667, \;\;
f/g=-1, \;\;
\Delta V_s=0, 
\label{potlam}
\end{equation}
and for $\Sigma$:
\begin{equation}
\alpha_s=0.616, \;\;
\alpha_v=0.667, \;\;
f/g=+1, \;\;
\Delta V_s=20 \;\mbox{MeV}. 
\label{potsig}
\end{equation}
We will study sensitivities of the reaction to deviations from 
these parameters.

\begin{figure}[p]
\centerline{\psfig{file=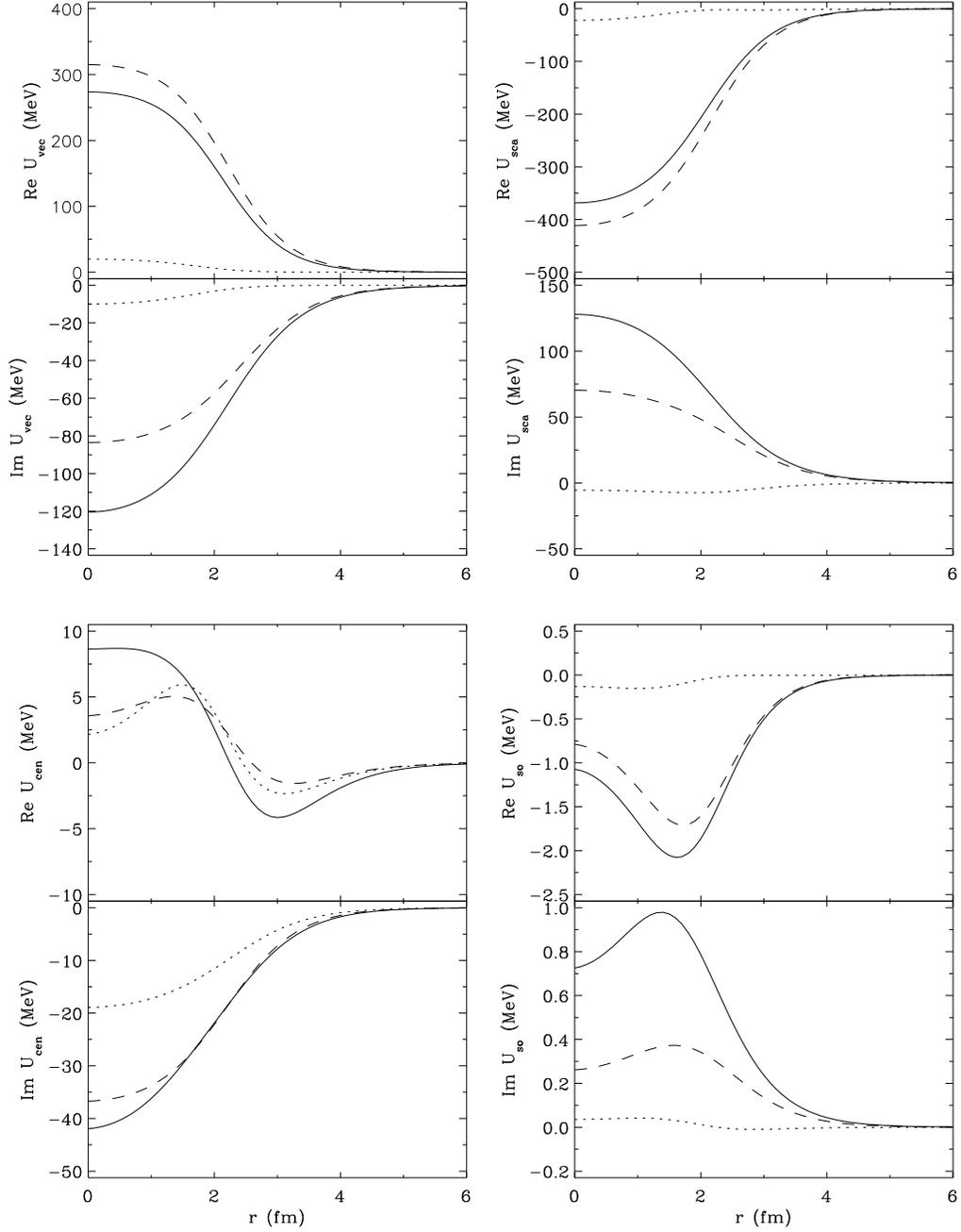,width=6in}}
\caption{Hyperon optical potentials for $^{12}$C at 300 MeV.
The upper panels show the Dirac vector and scalar potentials, while
the lower panels  show the corresponding Schr\"{o}dinger equivalent 
central and spin-orbit potentials.
The three curves correspond to the $\Lambda$ (dashed), $\Sigma^0$ (dotted), 
and proton (solid) potentials.}
\label{yopt-c300}
\end{figure}

\begin{figure}[p]
\centerline{\psfig{file=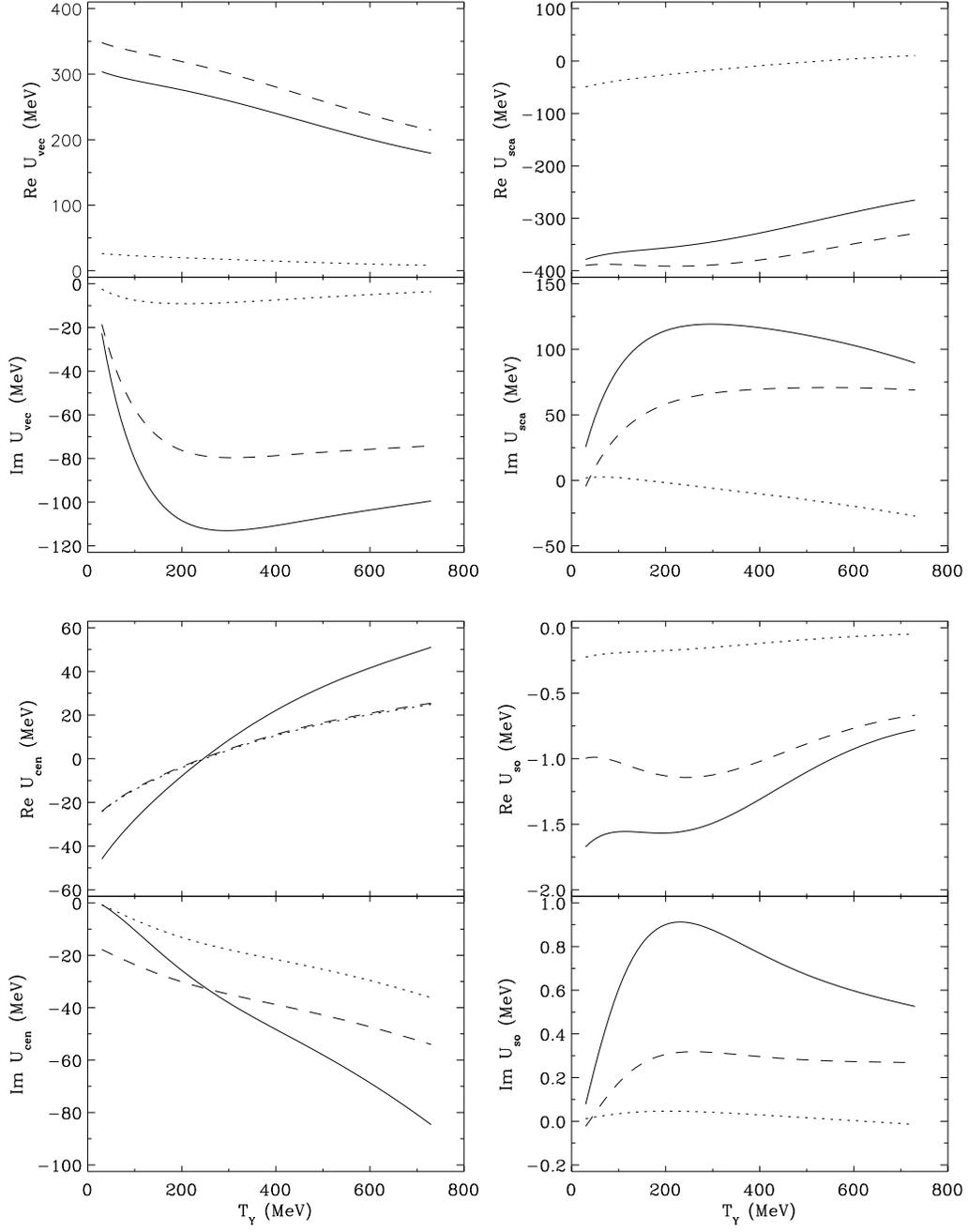,width=6in}}
\caption{Same as Fig.~\protect\ref{yopt-c300}, 
showing the energy dependence at a fixed distance of $r=1$ fm.}
\label{yopte-c}
\end{figure}

We generated hyperon distorted wavefunctions using the Schr\"{o}dinger 
equivalent potentials which have a central and a spin-orbit part:
$U(r)=U_{\rm cen}(r)+U_{\rm so}(r)\; {\mbox{\boldmath ${s}$}} 
\cdot {\mbox{\boldmath ${l}$}}$.
Note that the total spin-orbit part depends on 
the partial wave under consideration.
To get some idea about the hyperon potentials as compared to that of 
the nucleon, we show in Fig.~\ref{yopt-c300} a plot of the 
original vector and scalar potentials, and the corresponding 
Schr\"{o}dinger equivalent central and spin-orbit potentials
on $^{12}$C at 300 MeV, for $\Lambda$, $\Sigma^0$, and the proton. 
As expected, the hyperon potentials are weaker than that of the proton, 
and the $\Lambda$ potential is weaker than that of the $\Sigma$, 
especially for the spin-orbit part.
Fig.~\ref{yopte-c} shows a similar plot for the energy dependence
of the potentials at fixed distance $r=1$ fm.
The energy dependence is smooth.
The central potentials slowly increase with energy,
while the spin-orbit ones are relatively energy-independent.
The dependence is essentially the same for both $\Lambda$ and  $\Sigma$. 
Note that the central and spin-orbit potentials develop different 
energy dependence from that in the vector and scalar potentials. 
This can be traced to the energy-dependent factors in the nonrelativistic
reduction procedure.

\section{Results and Discussion}
\label{res}

As noted above, there is a great deal of kinematic flexibility in the 
reaction A$(\gamma,KY)$B. We decided to present our calculations under 
two kinematic arrangements: quasifree kinematics (small and fixed 
momentum transfer magnitude, $p_m$) and open kinematics (large variation 
of the momentum transfer). We will limit ourselves to coplanar setups 
with the hyperon on the opposite side of the kaon ($\phi_Y=180^\circ$). 
Such setups generally result in larger cross sections than out-of-plane 
setups. We will use $^{12}$C as an example, but our framework can easily 
be extended to other nuclei.
Since all possible channels can be explored if the reaction is 
measured exclusively on nuclei,  we try to provide as thorough 
an overview as possible 
by presenting results for all six channels.

\subsection{Quasifree Kinematics}
\label{res_qfree}

This setup is achieved by solving Eq.~(\ref{mcon}) and Eq.~(\ref{econ})
at fixed $E_\gamma$, $|{\mbox{\boldmath ${p}$}}_m|$ and $\theta_K$.
The quasifree kinematics closely resembles the two-body kinematics 
in free space, except here the reaction occurs 
on a bound nucleon with momentum $p_m$.
%The energy partitions are very close to those in free space 
%(with small influences from nuclear binding and recoil).
The hyperon angle will be shifted from its free space value 
by a certain amount depending on the value of $p_m$.
This kinematic arrangement has the feature that the energies of the 
outgoing particles vary in the whole angular range, making it maximally 
dependent on the the final state interactions and minimally sensitive to 
the details of the nuclear wavefunction.  The invariant mass of the outgoing
pair, denoted by $W$, stays within a narrow range.

In the following, we will present kaon angular distributions of the 
observables for the reactions 
$^{12}{\rm C}(\gamma,KY) ^{11}{\rm B}_{\rm g.s.}$ 
(the final nucleus is left in its ground state)
at $E_\gamma$=1.4 GeV and $p_m$=120 MeV. This value of $p_m$ yields maximal 
counting rates for $p$-shell nuclei.
For values of $\theta_K = 0^\circ, 30^\circ, 60^\circ, 90^\circ$,
the corresponding solutions are approximately,
$T_K=680, 474, 179, 42$ MeV, 
$T_Y=49, 255, 550, 687$ MeV, 
$\theta_Y = 21^\circ, 40^\circ, 24^\circ, 13^\circ $, and
$W=1885, 1921, 1902, 1886$ MeV for the $K^+\Lambda$ channel;
and
$T_K=565, 375, 107, 4$ MeV,
$T_Y=87, 277, 546, 648$ MeV,
$\theta_Y=15^\circ, 33^\circ, 19^\circ, 8^\circ $, and
$W=1881, 1912, 1894, 1877$ MeV for the $K^+\Sigma^0$ channel.
These energy ranges are well covered by the optical potentials.

Fig.~\ref{qfree-e14-dw} shows the effects of final state interactions.
Four different levels of approximations are shown for the coincidence 
cross section ($d^3\sigma$), the photon asymmetry ($A_\gamma$), and 
the hyperon recoil polarization ($A_Y$):
in Plane Wave Impulse Approximation (PWIA) where plane waves were used 
for the outgoing kaon and hyperon, in DWIA with hyperon FSI turned off, 
in DWIA with kaon FSI turned off, and in full DWIA.
Clearly, the angular distributions are peaked in the forward directions.
The magnitudes of the asymmetries $A_\gamma$ and $A_Y$ are sizeable 
and should be measurable in experiments.  
Our PWIA results agree qualitatively with the results of ref.
\cite{piekarewicz99}, the differences are be attributed to their use
of an older elementary amplitude.  As pointed out in that study,
the polarization observables especially can change widely with
different elementary operators.

The kaon FSI alone causes small reductions 
in the cross sections (about 10\%), and has little influence on the 
polarization observables.
The hyperon FSI alone causes larger reductions in 
the cross sections for the $K\Sigma$ channels (up to 40\%) than
for the $K\Lambda$ channels (up to 20\%).
Such behavior in the cross sections is consistent with our expectation 
since the $\Sigma$ 
potentials are stronger than the $\Lambda$ ones by construction.
What is interesting to observe is the interference of the two FSIs when 
both are turned on simultaneously. 
In the $K\Lambda$ channels the kaon and hyperon distortions appear to 
 combine with a small amount of destructive interference.  
However, in  the $K\Sigma$ channels,
 the two final state interactions 
constructively interfere in a way producing a DWIA cross section
that is enhanced compared to the one with only
 the hyperon FSI present.  
Thus, the kaon and hyperon distortions interfere with each
other in a complicated pattern, making the
extraction of the hyperon-nucleus potential more difficult.
This influence of the kaon FSI is also observed in the 
polarization observables.
As a result, the net effects of the FSIs on the cross sections 
are comparable in all six channels. 
We also point out that  $A_\gamma$
is more strongly affected by the FSIs in the $K\Sigma$ channels, 
especially $K^+\Sigma^-$, 
while it has little effect in the $K\Lambda$ channels.
However, the effects may be too small to be detected 
experimentally since the cross sections in the regions of 
large effects are rather small.

Fig.~\ref{qfree-e14-br} shows the individual contributions 
from the Born and resonance terms in the elementary production operator.
The calculations were performed in full DWIA.
As discussed in the previous section, the elementary
production process is resonance dominated. This fact is reflected in the 
angular distributions for quasifree production, which are almost
totally given by the resonant terms.
The photon asymmetry, on the other hand, displays some
significant interference patterns between Born and resonance 
contributions. The hyperon polarization is solely caused by
resonances since it samples only the imaginary part of the
elementary amplitude.
As expected, we find the relative contributions of Born
and resonance terms to depend only on W, rather than 
the momenta of the exiting particles.

It is clear that the three ingredients in the reaction
[see Eq.~(\ref{3d})], 
the elementary production process, the kaon FSI, and the hyperon FSI,
interfere coherently in a complicated fashion. 
It is reasonable to expect the interference 
to depend on the kinematics selected.
To study this possibility in the interest of searching for larger hyperon 
FSI effects, next we consider a different kinematic setup 
where the kaon energy is kept fixed.

\subsection{Open Kinematics}
\label{res_open}

This setup is achieved by solving Eq.~(\ref{mcon}) and Eq.~(\ref{econ})
for $E_Y$ and ${\mbox{\boldmath ${p}$}}_m$ at 
fixed angles, $E_K$ and $E_\gamma$.
The word `open' refers to the fact that the missing momentum 
${\mbox{\boldmath ${p}$}}_m$ is free to vary.
We will present observables as a function of the photon energy
for the same reactions $^{12}{\rm C}(\gamma,KY) ^{11}{\rm B}_{\rm g.s.}$
at $\theta_K=30^\circ$, $\theta_Y=35^\circ$, and $T_K=450$ MeV.
This is equivalent to  having a hyperon energy distribution 
according to Eq.~(\ref{econ}).
At the same time, it maps out the momentum distribution of the 
struck nucleon, and sweeps through the resonance region as indicated 
by the invariant mass $W$.
For values of $E_\gamma =1.3, 1.4, 1.5, 1.6$ GeV,
the corresponding solutions are approximately,
$T_Y =164, 264, 364, 464$ MeV, 
$p_m =101, 74, 155, 239$ MeV, and  
$W =1865, 1882, 1903, 1926$ MeV for the $K^+\Lambda$ channel;
and
$T_Y =87, 187, 289, 387$ MeV, 
$p_m =263, 136, 132, 200$ MeV, and  
$W =1941, 1949, 1965, 1986$ MeV for the $K^+\Sigma^0$ channel.

\begin{figure}
\centerline{\psfig{file=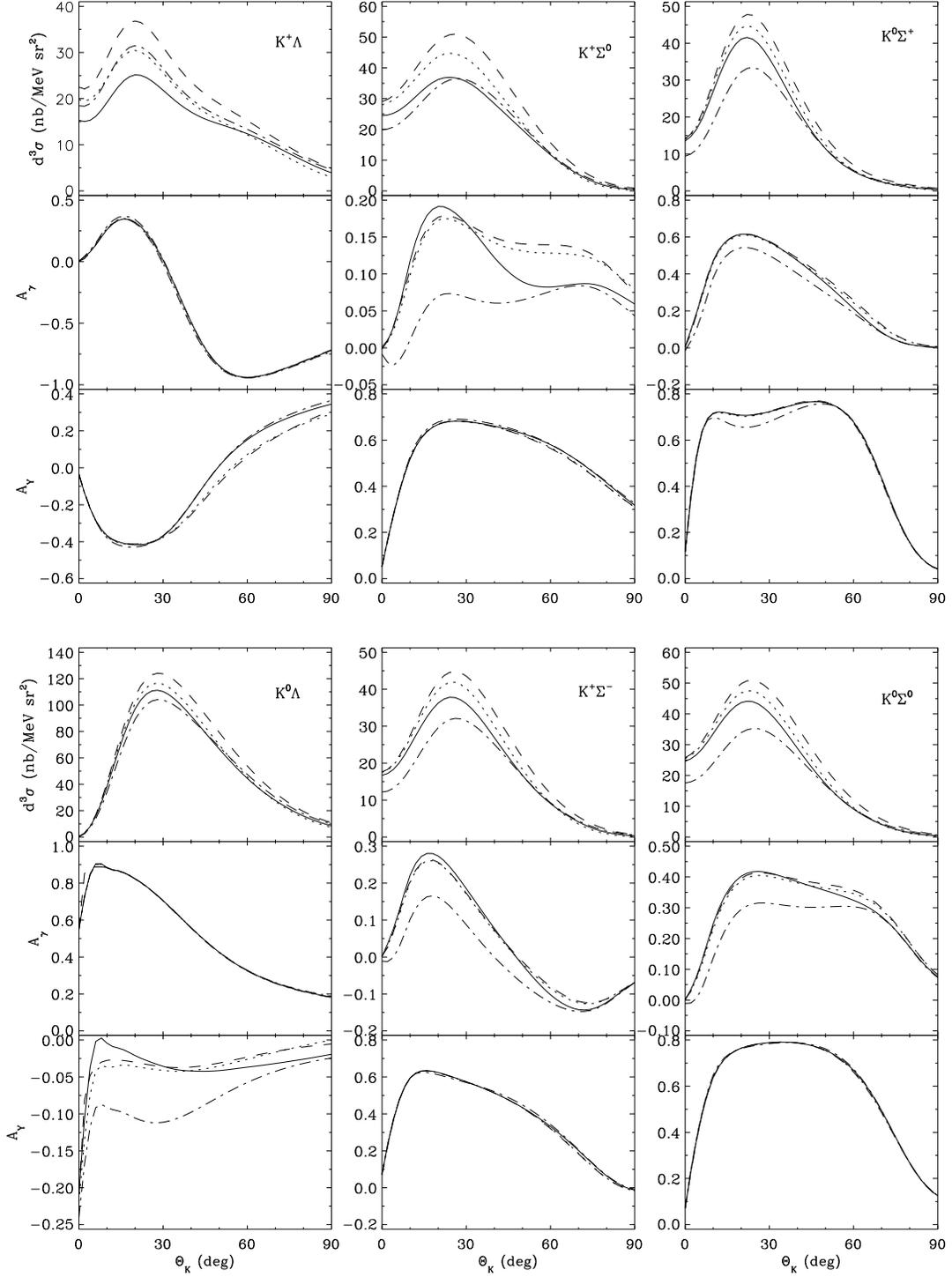,width=6in}}
\caption{Effects of final state interactions under quasifree kinematics
for the reaction $^{12}{\rm C}(\gamma,KY) ^{11}{\rm B}_{\rm g.s.}$
at $E_\gamma =1.4$ GeV and $p_m =120$ MeV.
The four curves correspond to calculations in PWIA (dashed), 
in DWIA with only kaon FSI (dotted), 
with only hyperon FSI (dash-dotted), and the full DWIA(solid).}
\label{qfree-e14-dw}
\end{figure}

\begin{figure}
\centerline{\psfig{file=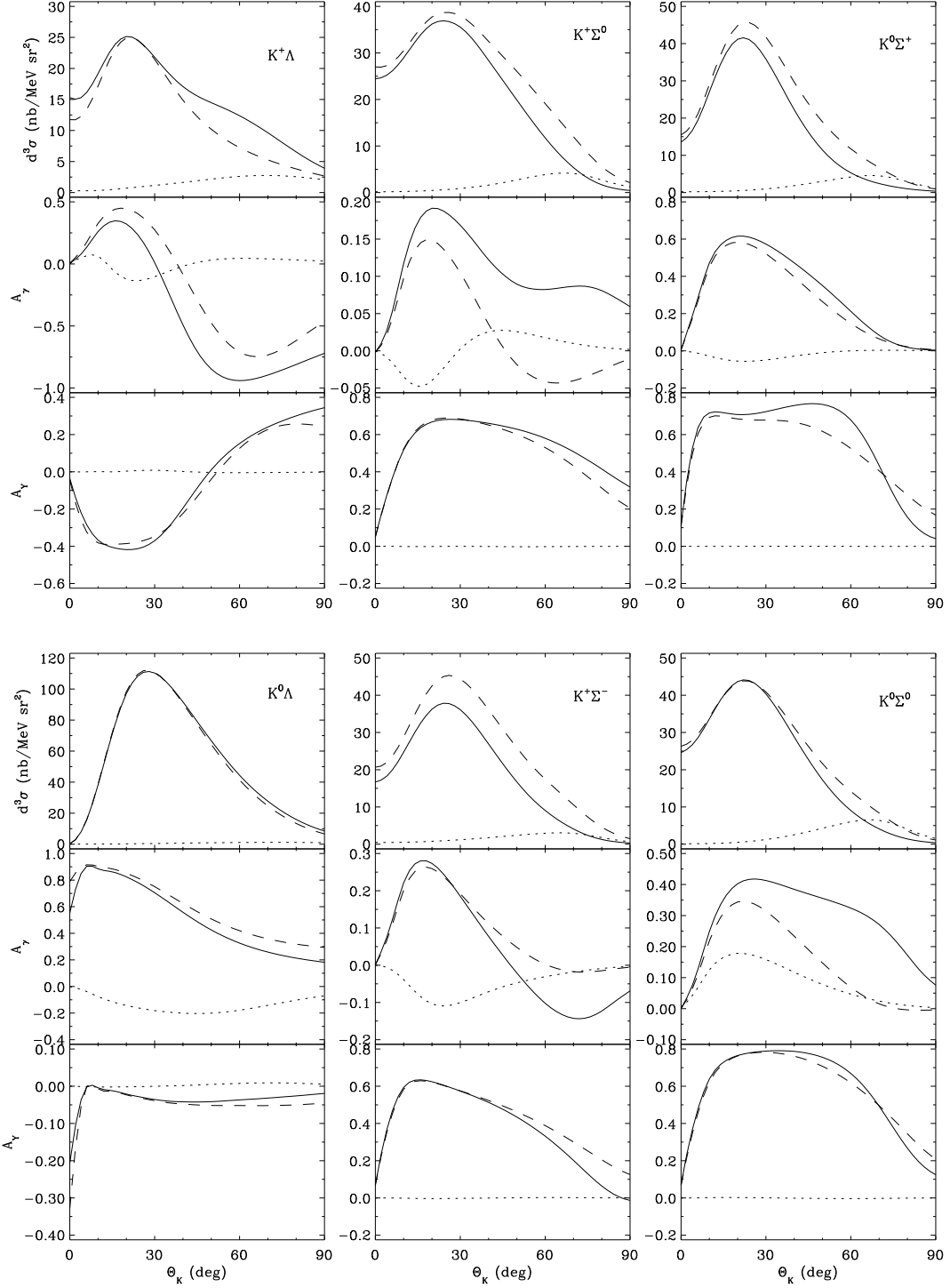,width=6in}}
\caption{Total and individual contributions from the Born and 
resonance terms under the quasifree kinematics of
Fig.~\protect\ref{qfree-e14-dw}.
The three curves correspond to the full (solid), 
Born only (dotted), and resonance only (dashed) contributions of the 
elementary amplitude.  The calculations were done in DWIA.}
\label{qfree-e14-br}
\end{figure}

\begin{figure}
\centerline{\psfig{file=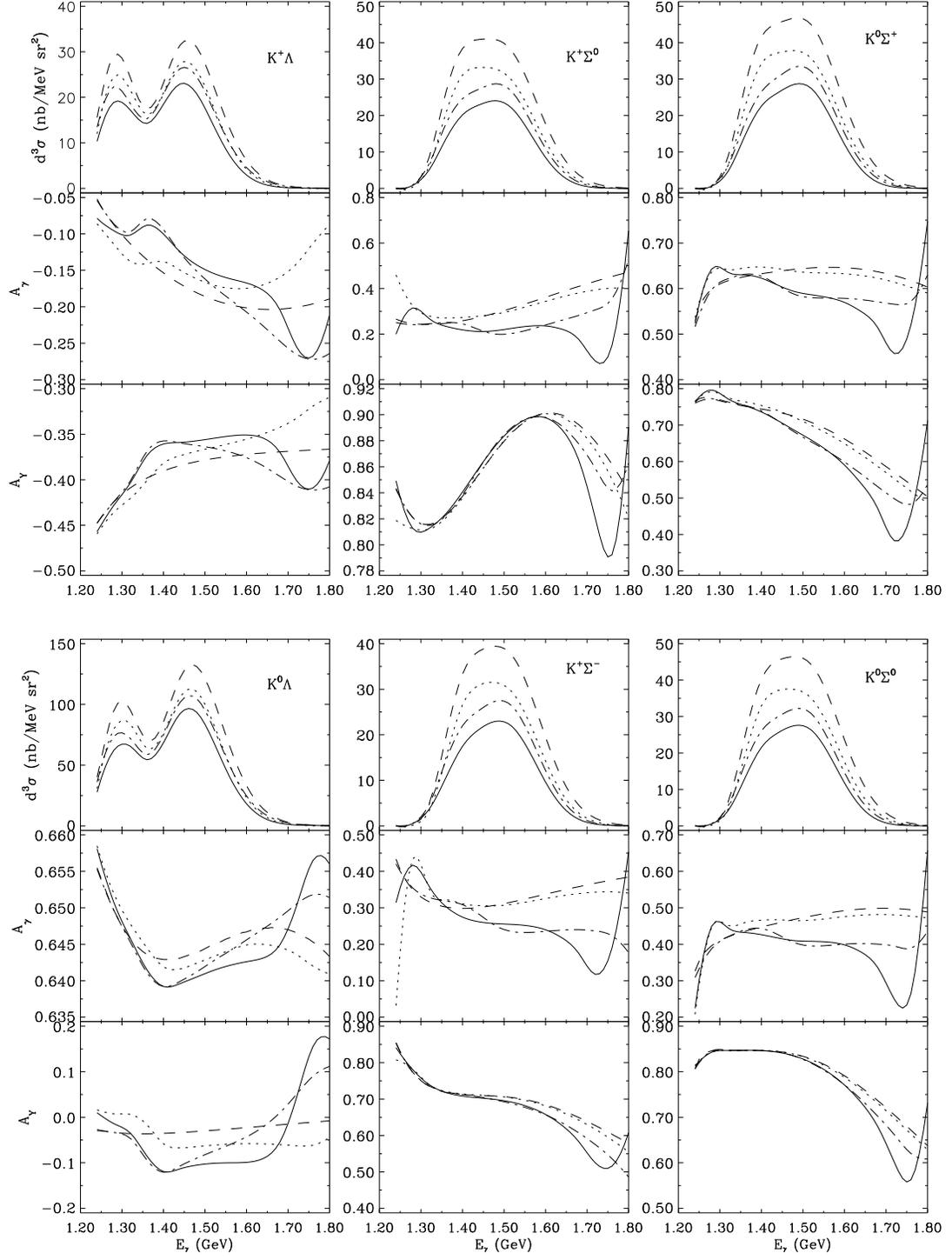,width=6in}}
\caption{Same as Fig.~\protect\ref{qfree-e14-dw}, 
but under open kinematics, with $\theta_K=30^\circ$, $\theta_Y=35^\circ$, 
and $T_K =450$ MeV.}
\label{open-egam-dw}
\end{figure}

Fig.~\ref{open-egam-dw} shows the effects of final state interactions
under this set of kinematics.
Inclusion of the kaon and hyperon FSI
leads to reductions of the cross sections up to a factor of two.
In most cases, FSI significantly affects the shape of the polarization 
observables. This clearly indicates that our finding of 
Fig.~\ref{qfree-e14-dw}, namely that most polarization observables are
independent of FSI, only holds true for selected kinematic situations. 
Thus, plane wave results as those presented
 in Ref.\cite{piekarewicz99} have to be treated with caution.
The conclusions obtained from Fig.~\ref{qfree-e14-dw} 
about the relative contributions of the FSIs to the cross sections remain 
true. But the role of the kaon FSI is now different as compared to quasifree 
kinematics;
it interferes constructively with the hyperon FSI in almost in all cases.
The double peaks in the cross section of the two $\Lambda$ channels 
are of kinematic origin; they come from the range
of values of $p_m$, which crosses the maximum of the 
$p$-shell single particle wavefunctions twice.

Having identified kinematic regions where large hyperon 
FSI effects are present, we now proceed to study 
the sensitivity of the observables
to the hyperon potential parameters, 
as given in Eq.~(\ref{potlam}) and Eq.~(\ref{potsig}).
In particular, we investigate which part of the hyperon-nucleus
optical potential can be studied best with quasifree kaon photoproduction
on nuclei.

\begin{figure}
\centerline{\psfig{file=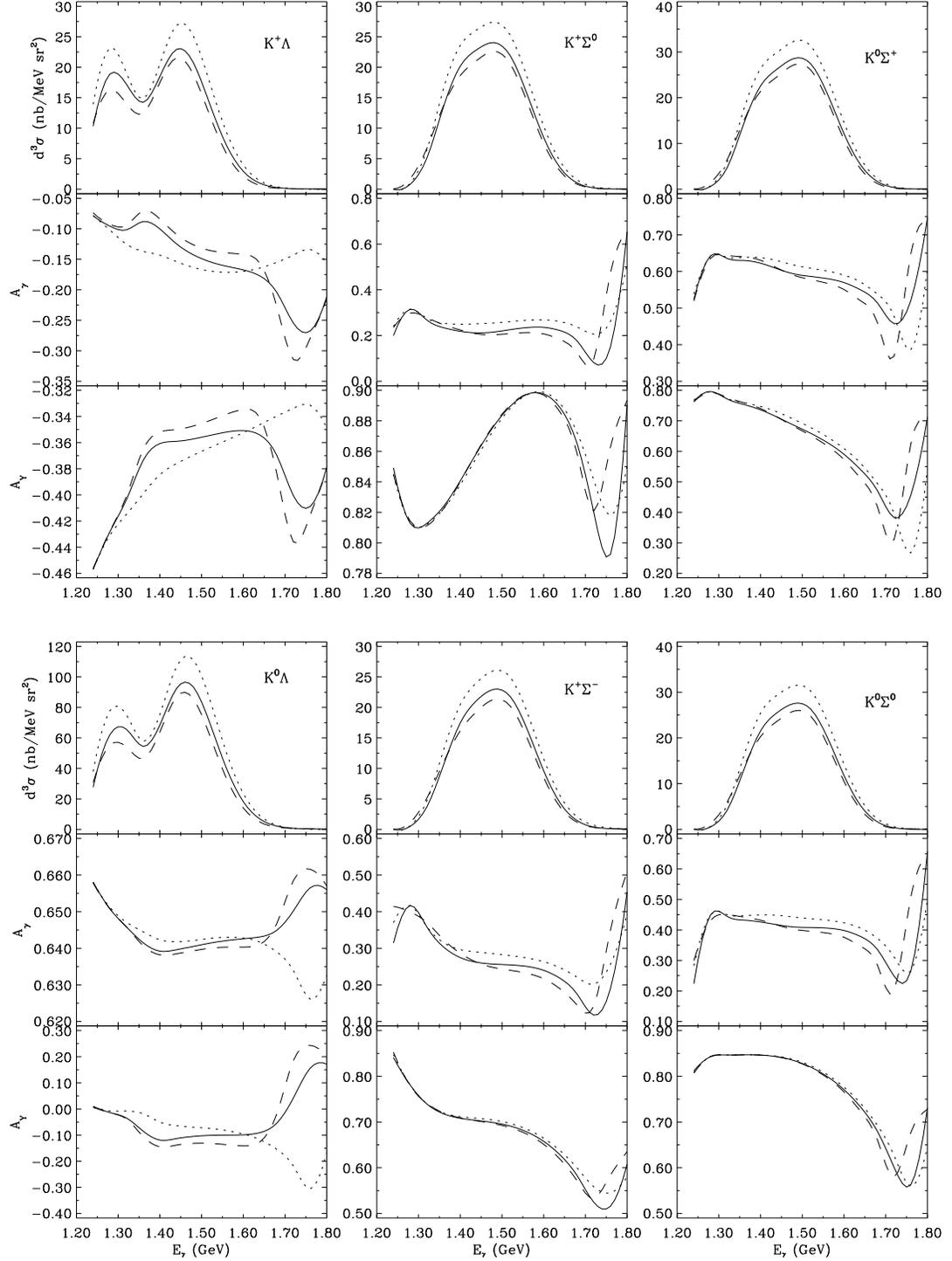,width=6in}}
\caption{Effects of varying the scaling factors 
in the hyperon potentials under  open kinematics. 
The solid line is in full DWIA ($\alpha_v\simeq\alpha_s\simeq 0.67$),
the dotted line is with $\alpha_v=0.333$ and $\alpha_s=0.345$,
and the dashed line is with $\alpha_v=\alpha_s=1$.}
\label{open-egam-alf}
\end{figure}

We varied the potential parameters $\alpha_v$ and $\alpha_s$ 
in order to modify the overall strength of the optical potentials.
Calculations were performed for two extreme cases, namely, 
reducing them by half in one case and setting them equal to one in the other. 
This corresponds to weakening and strengthening of the potentials, 
respectively. The results are shown in Fig.~\ref{open-egam-alf}.  
As expected, varying the strength of the hyperon potentials changes the cross 
sections by roughly scaling it up or down.
The polarization observables in the $K\Lambda$ channels are strongly 
modified, however, the biggest effects are found are higher energies
where the cross sections are small and more difficult to measure.
The asymmetries in the $K\Sigma$ channels also display moderate
sensitivities at higher energies but are generally less affected.

\begin{figure}
\centerline{\psfig{file=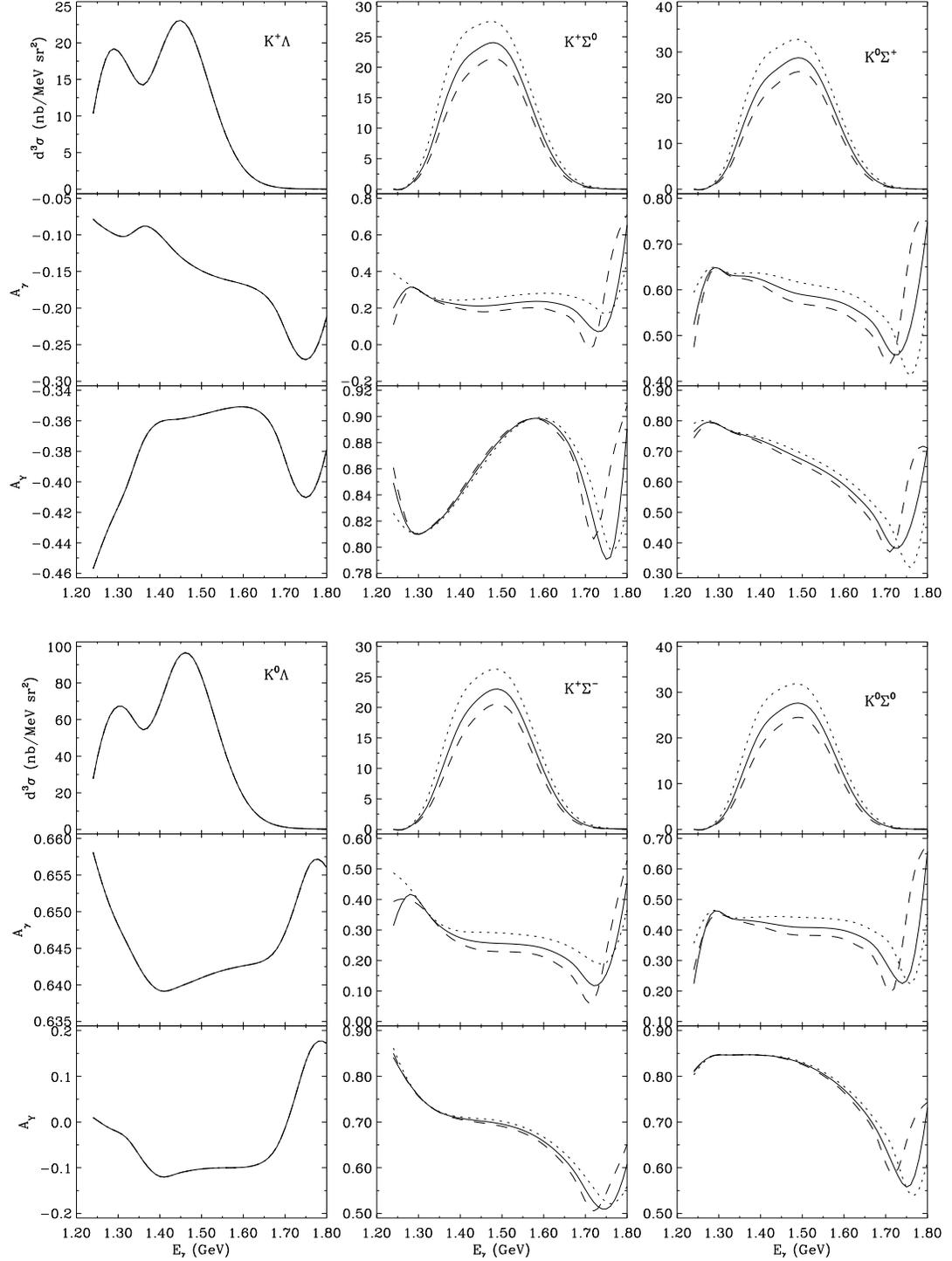,width=6in}}
\caption{Effects of varying the 
$\Sigma N \rightarrow \Lambda N$ conversion factors 
in the hyperon potentials under  open kinematics. 
The solid line displays the full DWIA calculation,
the dotted line shows the result with the conversion potential turned off,
and the dashed line is obtained by doubling the strength of the conversion 
potential.}
\label{open-egam-ass}
\end{figure}

Next, we varied the parameter $\Delta V_s$ which accounts for the 
$\Sigma N \rightarrow \Lambda N$ conversion in the $\Sigma$-nucleus
potential.  This conversion is known to be very important in few-body
hypernuclei, i.e., it leads to the binding of the hyper-triton and 
the correct energy spectrum in the A=4 systems.
Fig.~\ref{open-egam-ass} shows that the effects 
of either turning the conversion potential off or doubling its magnitude
on the observables for the $K\Sigma$ channels 
are essentially the same as the ones found from varying the overall 
strengths of the potentials.
This suggests that the two effects cannot be separated.
In this context, we also examined the sensitivity
to the central and spin-orbit parts
of the hyperon-nucleus potentials.
It turns out that, even for the polarization observables,
the hyperon FSI effects are 
almost entirely due to the central potentials. 
We furthermore investigated the sensitivity
to the tensor coupling terms  that were added  to 
the hyperon potentials (not shown).
We found again that in kinematic regions of appreciable cross section 
none of our observables are sensitive to the tensor coupling in any of the 
channels.

\begin{figure}
\centerline{\psfig{file=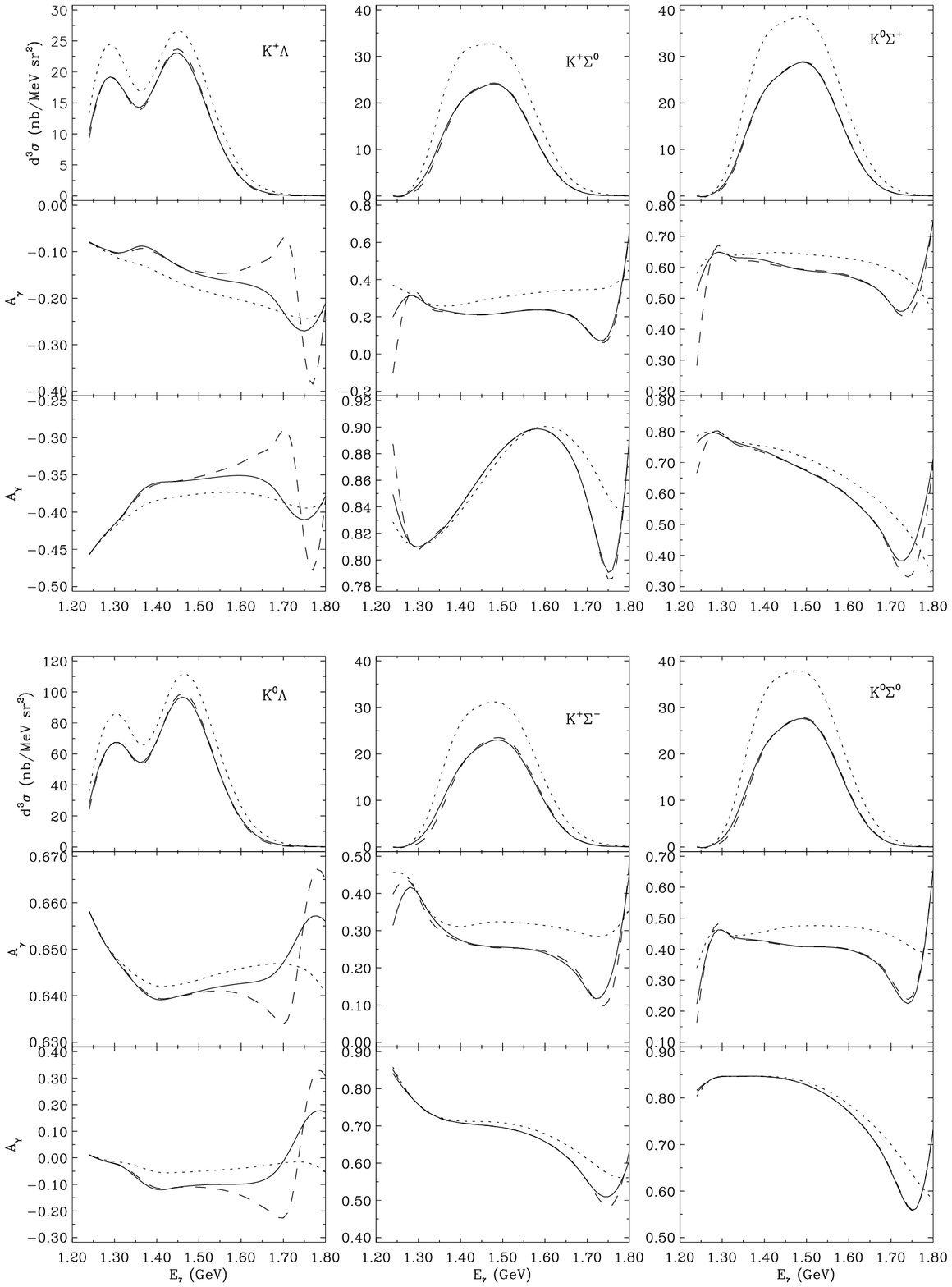,width=6in}}
\caption{Effects of turning on and off the real and imaginary parts
of the hyperon potentials under  open kinematics. 
The solid line is in full DWIA,
the dotted line is with the imaginary part turned off,
and the dashed line is with the real part turned off.}
\label{open-egam-ri}
\end{figure}

Finally, Fig.~\ref{open-egam-ri} displays the sensitivity of
the different observables to the real and imaginary parts
of the optical potentials. The reduction in the cross section is
caused solely by the imaginary parts, the real parts have almost no 
influence on the angular distributions. This should not come as a
surprise since the imaginary part of the potential removes flux 
from the matrix element and therefore leads to a reduction in the
cross sections.  In the model for the hyperon-nucleus
potentials adopted here, the parameters for the
real and imaginary parts of the potential are related,
this, however, needs not be true for more
sophisticated potentials developed in the future.
The situation is different for
the polarization observables, for the  $K\Lambda$ channels
both asymmetries show significant effects from the real part of the potential
at higher energies,
while for the $K\Sigma$ channels such effects can be found near threshold.
However, as before these are regions with very small cross sections, making 
a detailed study difficult.

\section{Conclusion} 
\label{con}

We have investigated the potential of the quasifree reactions 
A$(\gamma, K Y)$B to extract information on the 
hyperon-nucleus interaction through final state interactions.
Large differences were found between PWIA and DWIA results,
indicating the importance of both kaon and hyperon final state
interactions.  However, studying the hyperon-nucleus potentials
in detail is more difficult and can only be accomplished under
selected kinematics.

Several ingredients for this reaction
have to be known more precisely before
any quantitative conclusions about the hyperon-nucleus
potential could be drawn.
The most important is clearly the elementary
operator; while much progress has been made in
the last couple of years, both experimentally and theoretically,
more work must be done to gain a more precise understanding 
of the underlying dynamics. This is especially true
for the different $K \Sigma$ channels.
The $K^+$-nucleus interaction has been studied 
in great detail in the last decade; sophisticated
descriptions are available that can reproduce $K^+$-nucleus
elastic scattering data. 
The kaon FSI, despite being relatively weak in strength, 
plays a nontrivial role: 
It can interfere with the hyperon FSI to reduce or enhance the
combined FSI effects.  Future studies of this reaction
should therefore include improved kaon wave functions. 

The situation here is to be contrasted with the
experimentally very difficult, direct process of 
elastic scattering of hyperons off nuclear targets,
whose observables have been shown to display more substantial sensitivities 
to the hyperon potential in the calculations of Ref.~\cite{Cooper94}. 
Precise measurements of the 
quasifree kaon production process, complemented with
direct scattering wherever possible,
should enhance our understanding of the $Y$-nucleus 
interaction in the future.

The difficulty of extracting details on the hyperon-nucleus optical potentials 
can be turned into an advantage: For quasifree kinematics most polarization 
observables are not affected by either kaon or hyperon distortion effects;
thus a PWIA approach can be used to compare with experiment. For the $K \Lambda$
channels the photon asymmetry turns out to be the observable insensitive to
distortion while for the $K \Sigma$ channels it is the hyperon recoil polarization.
As suggested in ref.\cite{piekarewicz99} these observables may now be used 
to search for medium modifications of the elementary amplitude.  Especially
the formation, propagation and decay of higher-lying $N^*$ resonances may
be modified in the nuclear medium. Polarization observables free of distortion
would constitute an ideal tool to uncover such effects in exclusive channels.

\acknowledgements
We would like to thank E. Cooper for helpful communications on the 
hyperon potential. T.M. thanks the Center for Nuclear Studies for
the warm hospitality extended to him during his stay in Washington,
D.C. This work was supported in part by U.S. DOE under 
Grants DE-FG02-95ER-40907 (F.L., C.B., H.H.), DE-FG02-87ER-40370 (L.W.), and 
the University Research for Graduate Education (URGE) grant (T.M.).

%%%%%%%%%%%%%%%%%%%%%%%%% TABLES %%%%%%%%%%%%%%%%%%%%%%%%%%%%%%%%

\begin{table}[htbp]
\begin{center}
\caption{Helicity amplitudes for $N^* \rightarrow N + \gamma$ 
         \protect\cite{pdg98} and the ratio 
         of the neutral and charged coupling strengths. In the latter, 
         error-bars are not shown.}
\label{tab:helicity}
\renewcommand{\arraystretch}{1.4}
\begin{tabular}{lrrr}
Resonance & $S_{11}(1650)$ & $P_{11}(1710)$ & $P_{13}(1720)$ \\
[0.5ex]
\tableline
$J^\pi$&$ \frac{1}{2}^-$& $\frac{1}{2}^+$& $\frac{3}{2}^+$ \\
$A_{1/2}^p$ ($10^{-3}$ GeV$^{-1/2}$)&$53\pm 16$ & $ 9\pm 22$ & $-18\pm 30$\\
$A_{1/2}^n$ ($10^{-3}$ GeV$^{-1/2}$) & 
        $-15\pm 21$   & $-2\pm 14$  &  $1\pm 15$ \\
$A_{3/2}^p$ ($10^{-3}$ GeV$^{-1/2}$) & - & - & $-19\pm 20$     \\
$A_{3/2}^n$ ($10^{-3}$ GeV$^{-1/2}$) & - & - & $-29\pm 61$    \\
$g_{N^{*0}n\gamma}/g_{N^{*+}p\gamma}$&$-0.28$   &$-0.22  $ &-\\
$g_{N^{*0}n\gamma}^{(1)}/g_{N^{*+}p\gamma}^{(1)}$ & - &-&$-2.24$\\
$g_{N^{*0}n\gamma}^{(2)}/g_{N^{*+}p\gamma}^{(2)}$ & - & - & $+0.42$ \\
[0.5ex]
\end{tabular}
\end{center}
\end{table}

\begin{table}[htbp]
\begin{center}
\caption{Relative branching ratios $(x_i)$ for $S_{11}(1650)$, $P_{11}(1710)$, 
         and $P_{13}(1720)$ \protect\cite{feuster99}.}
\label{tab:branching}
\renewcommand{\arraystretch}{1.4}
\begin{tabular}{lcccc}
Resonance & $\pi N$ & $\pi\pi N$ & $\eta N$ & $K\Lambda$\\
[0.5ex]
\tableline
$S_{11}(1650)$ & 0.73 & 0.22 & 0.00 & 0.05 \\
$P_{11}(1710)$ & 0.00 & 0.51 & 0.32 & 0.17 \\
$P_{13}(1720)$ & 0.21 & 0.75 & 0.04 & 0.01 \\
[0.5ex]
\end{tabular}
\end{center}
\end{table}

\begin{table}[htbp]
\begin{center}
\caption{Extracted coupling constants in our models. Set I comes from our
        previous model which fits old photo- and electroproduction data
        \protect\cite{ben:mart_s96}, set II shows the result of our present
        calculation. Except for the Born terms only the product of
        coupling constants can be extracted from the fit.}
\renewcommand{\arraystretch}{1.4}
\label{tab:cc}
\begin{tabular}{lrr}
Coupling constants & Set I & Set II\\
[0.5ex]
\tableline
$g_{K\Lambda N}/\sqrt{4\pi}$ & $-3.09\pm 0.08$ & $-3.80$  \\
$g_{K\Sigma N}/\sqrt{4\pi}$ & $1.23\pm 0.06$ & $1.20$   \\
$\Theta_{\rm hd}~ (^\circ)$ & -~~ & $108\pm 4$\\
$\Phi_{\rm hd}~ (^\circ)$ & -~~ & $90\pm 6$\\
$\Lambda_{1}$ (GeV) & $0.85\pm 0.02$ & $0.80\pm 0.01$\\
$\Lambda_{2}$ (GeV) & -~~ & $1.88\pm 0.11$\\
[1ex]
\hline
$K\Lambda$ coupling \\
[0.5ex]
\hline
$g_{K^*K\gamma}~g^V_{K^*\Lambda N}/{4\pi}$&$-0.19\pm 0.01$&
$-0.51\pm 0.01$ \\
$g_{K^*K\gamma}~g^T_{K^*\Lambda N}/{4\pi}$ &$-0.12 \pm 0.02$& 
$ 0.67\pm 0.07$ \\
$g_{K_1K\gamma}~g^V_{K_1\Lambda N}/{4\pi}$ &-& $ 0.06\pm 0.07$ \\
$g_{K_1K\gamma}~g^T_{K_1\Lambda N}/{4\pi}$ &-& $ 0.37\pm 0.21$ \\
$g_{N^*(1650)N\gamma}~g_{K\Lambda N^*(1650)}/\sqrt{4\pi}$ &
$-0.06 \pm 0.01$& $-0.13\pm 0.00$ \\
$g_{N^*(1710)N\gamma}~g_{K\Lambda N^*(1710)}/\sqrt{4\pi}$ &
$-0.07 \pm 0.02$& $-0.09\pm 0.01$ \\
$g^{(1)}_{N^*(1720)N\gamma}~g_{K\Lambda N^*(1720)}/\sqrt{4\pi}$ &
- & $0.06\pm 0.00$ \\
$g^{(2)}_{N^*(1720)N\gamma}~g_{K\Lambda N^*(1720)}/\sqrt{4\pi}$ &
-& $0.94\pm 0.02$ \\
[1ex]
\hline
$K\Sigma$ coupling \\
\hline
$g_{K^*K\gamma}~g^V_{K^*\Sigma N}/{4\pi}$ &$-0.08 \pm 0.01$& 
$-0.31\pm 0.01$\\
$g_{K^*K\gamma}~g^T_{K^*\Sigma N}/{4\pi}$ &$-0.08 \pm 0.02$& 
$-0.60\pm 0.02$\\
$g_{K_1K\gamma}~g^V_{K_1\Sigma N}/{4\pi}$ &- & $-0.40\pm 0.04$\\
$g_{K_1K\gamma}g^T_{K_1\Sigma N}/{4\pi}$ &- & $-1.71\pm 0.22$\\
$g_{N^*(1650)N\gamma}~g_{K\Sigma N^*(1650)}/\sqrt{4\pi}$ &
$-0.01 \pm 0.02$& $-0.04\pm 0.00$\\
$g_{N^*(1710)N\gamma}~g_{K\Sigma N^*(1710)}/\sqrt{4\pi}$ &
$2.10 \pm 0.10$& $0.08\pm 0.02$\\
$g_{\Delta(1900)N\gamma}~g_{K\Sigma\Delta(1900)}/\sqrt{4\pi}$ &
$0.23 \pm 0.02$&$0.10\pm 0.00$\\
$g_{\Delta(1910)N\gamma}~g_{K\Sigma\Delta(1910)}/\sqrt{4\pi}$ &
$-0.99 \pm 0.09$&$0.36\pm 0.02$\\
$g_{K_1^0K^0\gamma}~/~g_{K_1^+K^+\gamma}$ &- & $0.26\pm 0.21$ \\
[1ex]
\hline
$\chi^2/N$ &5.99 & 3.45  \\
[0.5ex]
\end{tabular}
\end{center}
\end{table}

%%%%%%%%%%%%%%%%%%%%%%%%%% FIGURES %%%%%%%%%%%%%%%%%%%%%%%%%%%%%%%%%%%


% ------------- open ---------------------

\begin{references}

\bibitem{Bennhold89} C. Bennhold and L. E. Wright,
 Phys. Rev. C {\bf 39}, 927 (1989);
 Phys. Lett. B {\bf 191}, 11 (1987); 
 Prog. Part. Nucl. Phys. {\bf 20}, 377 (1988).

\bibitem{Cotanch86} S. R. Cotanch and S. S. Hsiao, 
 Nucl. Phys. {\bf A450}, 419c (1986).

\bibitem{Rosenthal88} A. S. Rosenthal, D. Halderson, K. Hodgkinson,
 and F. Tabakin, 
 Ann. Phys. (NY) {\bf 184}, 33 (1988); 
 A. S. Rosenthal, D. Halderson, and F. Tabakin,
 Phys. Lett. B {\bf 182}, 143 (1986); 

\bibitem{Cohen89} J. Cohen, 
 Phys. Rev. C {\bf 32}, 543 (1985);
 Int. J. Mod. Phys. A {\bf 4}, 1 (1989).

\bibitem{kaplan98} D.B. Kaplan, M.J. Savage and M.B. Wise, 
Phys. Lett. B {\bf 424}, 390 (1998); Nucl. Phys. B {\bf 534} 329 (1998).

\bibitem{Bertini81} R. Bertini {\it et al.},
 Nucl. Phys. {\bf A368}, 365 (1981).

\bibitem{Chrien88} R. E. Chrien {\it et al.},
 Nucl. Phys. {\bf A478}, 705c (1988).

 \bibitem{piekarewicz99} L. J. Abu-Raddad and J. Piekarewicz, 
 Phys. Rev. C {\bf 61} 014604 (1999).

\bibitem{Bianchi93} 
N. Bianchi {\it et al.}, Phys. Lett. B {\bf 299}, 219 (1993);
{\bf 309}, 5 (1993); {\bf 325}, 333 (1993), 
Phys. Rev. C {\bf 54}, 1688 (1996);
M. Anghinolfi {\it et al., ibid.} {\bf 47}, R922 (1993);
M MacCormick {\it et al., ibid.} {\bf 55} 1033 (1997).

 \bibitem{Hyde91} C. E. Hyde-Wright (spokesperson), 
``Quasifree Strangeness Production in Nuclei'',
 CEBAF experiment 91-014.

\bibitem{Lee93} F. X. Lee, L. E. Wright, C. Bennhold,
 Phys. Rev. C {\bf 48}, 816 (1993);
 Phys. Rev. C {\bf 55}, 318 (1997).

\bibitem{Lee96} F. X. Lee, L. E. Wright, C. Bennhold, and L. Tiator,
 Nucl. Phys. {\bf A603}, 345 (1996).

\bibitem{Bennhold97} C. Bennhold, F. X. Lee, T. Mart, and L. E. Wright,
 Nucl. Phys. {\bf A639}, 227c (1998);
 nucl-th/9712075.


\bibitem{ben:saghai} J. C. David, C. Fayard, G. H. Lamot, and B. Saghai,
 Phys. Rev. C {\bf 53}, 2613 (1996).

\bibitem{ben:williams} R. A. Williams, C.-R. Ji, and S. R. Cotanch,
 Phys.\ Rev.\ C {\bf 46}, 1617 (1992).

\bibitem{ben:mart95} T. Mart, C. Bennhold, and C. E. Hyde-Wright, 
 Phys. Rev. C {\bf 51}, R1074 (1995). 

\bibitem{ben:mart_s96} C. Bennhold, T. Mart, and D. Kusno, in  
 {\it Proceedings of the CEBAF/INT  Workshop on $N^*$ Physics, Seattle,
 USA, 1996}, edited by T.-S. H. Lee and W. Roberts,
 (World Scientific, Singapore, 1997), p. 166.

\bibitem{feuster98}  T. Feuster and U. Mosel, 
 Phys. Rev. C {\bf 58}, 457 (1998).

\bibitem{feuster99}  T. Feuster and U. Mosel, 
 Phys. Rev. C {\bf 59}, 460 (1999).

\bibitem{han99}
B.\,S. Han, M.\,K. Cheoun, K.\,S. Kim, and I.-T. Cheon,
nucl-th/9912011.

\bibitem{dytman2000} T.P. Vrana, S.A. Dytman, and T.-S.H. Lee,
 Phys. Rep. {\bf 328}, 181 (2000).

\bibitem{manley92} D.M. Manley, E.M. Saleski, 
 Phys. Rev. D {\bf 45}, 4002 (1992).

\bibitem{saxon80} D. H. Saxon {\it et al.}, 
 Nucl. Phys. {\bf B162}, 522 (1980).

 \bibitem{bell83} K. W. Bell {\it et al.}, 
  Nucl. Phys. {\bf B222}, 389 (1983).
 
\bibitem{ben:adelseck} R. A. Adelseck and L. E. Wright, 
 Phys. Rev. C {\bf 38}, 1965 (1988).               

\bibitem{dennery} P. Dennery, 
 Phys. Rev. {\bf 124}, 2000 (1961).

\bibitem{thom} H. Thom, 
 Phys. Rev. {\bf 151}, 1322 (1966).

\bibitem{deo} B. B. Deo and A. K. Bisoi, 
 Phys. Rev. D {\bf 9}, 288 (1974).

\bibitem{abw85} R. A. Adelseck, C. Bennhold, and L. E. Wright, 
 Phys. Rev. C {\bf 32}, 1681 (1985).

\bibitem{adelseck90} R. A. Adelseck and B. Saghai, 
 Phys. Rev. C {\bf 42}, 108 (1990).

\bibitem{singer} P. Singer and G. A. Miller, 
 Phys. Rev. D {\bf 33}, 141 (1986).

\bibitem{zhenping} Zhenping Li, Hongxing Ye, and Minghui Lu, 
 Phys. Rev. C {\bf 56}, 1099 (1997).

% Included this reference here [hh]
\bibitem{kroll54} 
N. M. Kroll and M. A. Ruderman, 
Phys.\ Rev.\ C{\bf 93}, 233 (1954).

\bibitem{haberzettl97} H. Haberzettl, 
 Phys. Rev. C {\bf 56}, 2041 (1997). 

\bibitem{ben:ohta} K. Ohta, 
 Phys. Rev. C {\bf 40}, 1335 (1989).

\bibitem{hbmf98} H. Haberzettl, C. Bennhold, T. Mart, and T. Feuster, 
 Phys. Rev. C {\bf 58}, R40 (1998); 
H. Haberzettl, nuch-th/0003058.
%"Preserving the gauge invariance of meson production currents 
%in the presence of explicit final-state interactions".

\bibitem{saphir98} {\footnotesize SAPHIR} Collaboration: 
 M. Q. Tran {\it et al}., 
 Phys. Lett. B {\bf 445}, 20 (1998).

\bibitem{bockhorst} M. Bockhorst {\it et al}., 
 Z. Phys. C {\bf 63}, 37 (1994).

\bibitem{capstick98} S. Capstick and W. Roberts, 
 Phys. Rev. D {\bf 58}, 074011 (1998).

\bibitem{mart99} T. Mart and C. Bennhold, 
 Phys. Rev. C {\bf 61}, 012201(R) (1999).

\bibitem{guidal} M. Guidal, J.-M. Laget, and M. Vanderhaeghen, 
 Nucl. Phys. {\bf A627}, 645 (1997).

\bibitem{carlson2000} C.E. Carlson, hep-ph/0005169

\bibitem{elba98} C. Bennhold, T. Mart, A. Waluyo, H. Haberzettl, G. Penner,
        T. Feuster, and U. Mosel, in \textit{Proceedings of the Workshop on 
        Electron Nucleus Scattering}, Elba, Italy, 1998, edited by
        O. Bennhar, A. Fabrocini, and R. Schiavilla, Edizioni ETS, Pisa, 1999,
        p. 149, nucl-th/9901066.

\bibitem{ben:sibirtsev98} 
 A. Sibirtsev, K. Tsushima, W. Cassing, and A. W. Thomas,
 Nucl. Phys. {\bf A646}, 427 (1999).

\bibitem{waluyo2000} A. Waluyo {\it et al.}, Proc. of the 18th Indonesian
   National Symposium, Jakarta, Indonesia, April 24-26, 2000 (in press).

\bibitem{reinhard} R. Schumacher, private communication.

\bibitem{sanabria2000} F.J. Klein, J.C. Sanabria, and J.D. Kellie, 
co-spokespersons, "Photoproduction of Kaons
off Protons Using a Linearly Polarized Beam of Photons", JLab proposal in preparation.

\bibitem{kamalov} S. S. Kamalov, J. A. Oller, E. Oset, and M. J. 
 Vicente-Vacas, 
 Phys. Rev. C {\bf 55}, 2985 (1997).

\bibitem{Cooper94} E. D. Cooper, B. K. Jennings, and J. Mare\u{s},
Nucl. Phys. {\bf A580}, 419 (1994);
Nucl.. Phys. {\bf A585}, 157 (1995).

\bibitem{Cooper93} E. D. Cooper, S. Hama, B. C. Clark, and R. L. Mercer,
 Phys. Rev. C {\bf 47}, 297 (1993).

\bibitem{Mares94} J. Mare\u{s}, B. K. Jennings, and E. D. Cooper,
 Prog. Theor. Phys. (Suppl.) {\bf 117}, 415 (1994).

\bibitem{pdg98} C. Caso {\it et al}., 
 Eur. Phys. J. C {\bf 3}, 1 (1998).

\bibitem{old_data} Aachen-Berlin-Bonn-Hamburg-Heidelberg-M\"unchen
 Collaboration, 
 Phys. Rev. {\bf 188}, 2060 (1969). 
 A list of references for the old data can be found in 
 Ref.~\cite{saphir98}.

\bibitem{benn97} C. Bennhold {\it et al}., 
 Nucl. Phys. {\bf A639}, 209c (1998).

\end{references}
\end{document}